%% file: main.tex
\pdfoutput=1
\documentclass[aps,superscriptaddress,twocolumn,prd,nofootinbib,longbibliography,floatfix]{revtex4-2}

\usepackage{amsmath, amssymb, amsfonts, amsthm, latexsym, epsfig, mathrsfs, xcolor, bbm, slashed,braket,bm}

\usepackage[inline]{enumitem}

\usepackage{setspace}
\usepackage[marginal, multiple]{footmisc}

\usepackage[T1]{fontenc}
\usepackage[utf8]{inputenc}
\usepackage{lmodern}
\usepackage[normalem]{ulem}

\usepackage[unicode, colorlinks, allcolors=blue!70!black, linktocpage, pdfusetitle]{hyperref}
\usepackage[all]{hypcap}

\definecolor{indigo(dye)}{rgb}{0.0, 0.25, 0.42}
 
\numberwithin{equation}{section}

\usepackage{cleveref}

\usepackage{cancel}
\usepackage{tikz}
\usepackage{microtype}

\setlist[enumerate]{noitemsep, label=(\arabic*), ref=(\arabic*)}

\newlist{condlist}{enumerate}{2}
\setlist[condlist,1]{noitemsep, label=(\arabic*), ref=(\arabic*)}
\setlist[condlist,2]{noitemsep, label=(\alph*), ref=(\arabic{condlisti}.\alph*)}
\crefname{condlisti}{condition}{conditions}
\crefname{condlistii}{condition}{conditions}

\renewcommand\thesection{\arabic{section}}
\renewcommand\thesubsection{\arabic{subsection}}

\colorlet{mydarkblue}{blue!50!black}
\colorlet{mylightblue}{mydarkblue!6}
\colorlet{myvlightblue}{mydarkblue!3}
\colorlet{mypurple}{blue!40!red!80!black}
\colorlet{mydarkpurple}{blue!40!red!50!black}
\colorlet{mylightpurple}{mydarkpurple!80!red!6}
\makeatletter
\def\p@subsection{\thesection.}
\def\p@subsubsection{\thesection.\thesubsection.}
\makeatother 

\theoremstyle{plain}

\theoremstyle{definition}

\theoremstyle{remark}
\newtheorem{remark}{Remark}

\crefname{equation}{eq.}{eqs.}
\Crefname{equation}{Equation}{Equations}
\creflabelformat{equation}{#2#1#3}

\crefname{section}{sec.}{secs.}
\crefname{appendix}{appendix}{appendices}
\crefname{figure}{fig.}{figs.}

\crefname{definition}{Def.}{Defs.}
\crefname{prop}{Prop.}{Props.}
\crefname{lemma}{Lemma}{Lemmas}
\crefname{corollary}{Cor.}{Cors.}
\crefname{thm}{Theorem}{Theorems}
\crefname{remark}{Remark}{Remarks}

\crefname{ass}{Assumptions}{Assumptions}
\crefname{property}{Properties}{Properties}

\newcommand{\be}{\begin{equation}\begin{aligned}}
\newcommand{\ee}{\end{aligned}\end{equation}}

\newcommand{\lb}{\left}
\newcommand{\rb}{\right}

\newcommand{\mc}{\mathcal}

\newcommand{\ms}{\mathscr}
\newcommand{\mf}{\mathfrak}
\newcommand{\bb}{\mathbb}

\newcommand{\eqsp}{\, ,\quad}

\newcommand{\nfrac}[2]{{{}^#1\!\!/\!_#2}}

\newcommand{\half}{\nfrac{1}{2}}

\newcommand{\abs}[1]{\lb\vert\, #1 \,\rb\vert}
\newcommand{\norm}[1]{\lb\Vert\, #1 \,\rb\Vert}
\newcommand{\inp}[1]{\lb\langle #1 \rb\rangle}

\newcommand{\Lie}{\pounds}
\newcommand{\hatLie}{\Lie\kern-0.25em\hat{\vphantom{\Lie{}}}\kern0.25em}
\newcommand{\defn}{\mathrel{\mathop:}=} %shrtct for definition operator

\DeclareMathOperator{\sech}{sech}

\let\oldint\int
\renewcommand{\int}{\oldint\limits}

\newcommand{\op}[1]{\boldsymbol{#1}}
\newcommand{\1}{\op{1}}

\newcommand{\Alg}{\mathscr{A}}

\newcommand{\Hilb}{\mathscr{H}}

\newcommand{\Fock}{\mathscr{F}}

\newcommand{\EM}{\textrm{EM}}

\newcommand{\KG}{\textrm{KG}}

\newcommand{\GR}{\textrm{GR}}

\newcommand{\BD}{\textrm{BD}}

\newcommand{\aut}{\mf{a}}

\newcommand{\dS}{\textrm{dS}}

\newcommand{\hor}{\mathcal{H}}

\newcommand{\scri}{\ms I}

\newcommand{\inn}{\textrm{in}}

\newcommand{\NCM}{\textrm{SI}}

\begin{document}

\title{ Vacua and infrared radiation in de Sitter quantum field theory} 

\author{Jonah Kudler-Flam}
\email{jkudlerflam@ias.edu}
\affiliation{School of Natural Sciences, Institute for Advanced Study, Princeton, NJ 08540, USA}
\affiliation{Princeton Center for Theoretical Science, Princeton University, Princeton, NJ 08544, USA}

\author{Kartik Prabhu}
\email{kartikprabhu@rri.res.in}
\affiliation{Raman Research Institute, Sadashivanagar, Bengaluru 560080, India.} 

\author{Gautam Satishchandran}
\email{gautam.satish@princeton.edu}
\affiliation{Princeton Gravity Initiative, Princeton University, Princeton, NJ 08544, USA}

%\date{\today}

\begin{abstract}
We analyze the asymptotic behavior of quantum fields and perturbative quantum gravity in de Sitter space. We show that the necessary and sufficient condition for the existence of a de Sitter invariant vacuum state in the free theory is if the local field observables commute with the ``memory observable'' on any cosmological horizon. This criterion yields simple, gauge-invariant proofs of the existence of a de Sitter invariant vacuum for the (source-free) massive scalar field, electromagnetic field and linearized gravitational field. However, the massless, minimally coupled scalar does not satisfy this criterion and, consequently, there is no de Sitter invariant vacuum state. In this case, we construct the Hilbert space of normalizable states that have ``square integrable'' distributions in the memory and the conjugate constant mode. This Hilbert space has a unitary representation of the de Sitter symmetry group but no de Sitter invariant normalizable state. However even for free theories with a normalizable de Sitter invariant vacuum we show, by a simple example, that in the presence of a source a large number of infrared particles can be produced if the source interacts with the field for timescales much longer than the Hubble time. In the limit as the source persists forever it emits an infinite number of infrared particles. 

\end{abstract}

\maketitle

%%===================================================================================
\section{Introduction}
\label{sec:intro}

The current cosmological observations are consistent with a universe that initially underwent a period of rapid accelerated expansion known as ``inflation'' driven by the large potential energy of an inflaton field \cite{2023arXiv231213238E}. The quantum fluctuations of the inflaton give rise to density perturbations whose ``scale-free'' spectrum agrees precisely with high precision measurements of the cosmic microwave background (CMB) \cite{Planck:2018vyg}. This model, together with the $\Lambda$CDM model of cosmology successfully accounts for the formation of structure in our universe \cite{Bernardeau:2001qr}, the abundances of light elements \cite{Cyburt:2015mya} as well as the current accelerated expansion of our universe due to ``dark energy'' \cite{SupernovaCosmologyProject:1998vns,SupernovaSearchTeam:1998fmf}. 

During the inflationary era as well our current dark energy dominated era, the universe is well-described by a de Sitter spacetime and any quantum effects are treated perturbatively off of this background. At spacetime distances within a Hubble radius, the behavior of quantum fields have been shown to give important contributions such as the scalar and tensor fluctuations of the CMB induced by fluctations of the inflaton and linearized gravitons \cite{mukhanov1992theory,Weinberg:2008zzc} as well as small, non-Gaussianities \cite{Maldacena:2002vr}. However, at larger distances or later times, the general behavior of quantum fields and (perturbative) quantum gravity in de Sitter spacetime is much less understood. Due to the exponential expansion of the universe, one might expect that any quantum field will exponentially decay at sufficiently late times. The behavior of quantum fields at asymptotically early and late times in de Sitter has been investigated by many authors in the literature, and for various theories, see, e.g., \cite{Polyakov:2007mm,Polyakov:2009nq,Polyakov:2012uc,Krotov:2010ma,Anderson:2013ila,Ford:1984hs,Tsamis:1996qm,Tsamis:1996qq,Weinberg:2006ac,Mottola:2010gp,Anderson:2013zia,Mottola:1985qt,Mazur:1986et,Rouhani:2004zs,Gorbenko:2019rza}, with partly contradicting claims. Indeed, it has been argued by several authors that de Sitter symmetries may be ``broken'' in the free theory and/or in the interacting theory due to the existence and/or emission of an infinite number of ``infrared particles.'' However, due to issues such as gauge invariance and the complexity of perturbative calculations in de Sitter, the general validity, origin, and physical significance of these effects have been widely debated, see e.g., \cite{Tsamis:1996qq,Garriga:2007zk}. Indeed, for these reasons, even the existence or non-existence of a de Sitter invariant vacuum for the free graviton has been the subject of significant debate for the past 35 years, e.g., \cite{Ford:1977dj,Allen:1986dd,Higuchi:1986py,Antoniadis:1986sb,Tsamis:1993ub,Kleppe:1993fz,Higuchi:2000ye,Miao:2009hb,Urakawa:2010it,Higuchi:2011vw,Miao:2011fc,Mora:2012zi,Fewster:2012bj,Morrison:2013rqa,Bernar:2014lna,Frob:2014fqa,Woodard:2015kqa,Gerard:2024eef}. 

The purpose of this paper is to show that infrared behavior of quantum states is controlled by a simple gauge-invariant observable on the cosmological horizon: {\em the horizon memory}. In the usual approach, the generic early and late time behavior is obtained by propagating suitable initial data on a spacelike Cauchy surface. In this paper, we will instead consider the ``characteristic initial value problem'' of quantum fields where the initial data is specified on any cosmological horizon. In this approach, the asymptotic behavior at late affine times along any horizon is now a {\em kinematic} question of the allowed space of initial data as opposed to a dynamical question. This perspective will allow us to straightforwardly diagnose existence or non-existence of infrared radiation in the free theory and illustrate how such radiation can be produced in the presence of a ``source.'' We expect that the infrared effects that we discuss are related to many of the infrared effects discussed in the literature and believe our framework may be useful for resolving certain questions, though do not claim that it addresses all issues.

In general, the horizon memory is the difference in the value of the ``free data'' of the field at early and late times on any cosmological horizon. If $V$ is the affine parameter on the horizon and $x^{A}$ are spherical coordinates on the horizon cross-sections then, for any scalar field $\Phi$ on the horizon, the horizon scalar memory is 
\begin{equation}
\label{eq:memscalarintro}
\Delta(x^{A}) \defn \int_{-\infty}^{\infty}dV~\partial_{V}\Phi(V,x^{A}) = \Phi\big\vert_{V=\infty} - \Phi\big\vert_{V=-\infty}.
\end{equation}
For Maxwell fields, the free data on the horizon are the angular components $A_{A}(V,x^{A})$ on the horizon and the horizon electromagnetic memory is the integral 
\begin{equation}
\label{eq:EMmemintro}
\Delta_{A}(x^{A})\defn -\int_{-\infty}^{\infty}dV~E_{A}(V,x^{A}) = A_{A}\big\vert_{V=\infty} - A_{A}\big\vert_{V=-\infty}.
\end{equation}
where the electric field $E_{A}=-\partial_{V}A_{A}$ and we have chosen a gauge where $A_{V}=0$ on the horizon. For gravitational perturbations, the free data $\gamma_{AB}$ is the trace-free part of the metric perturbation on the horizon and the horizon gravitational memory is 
\begin{align}
\Delta_{AB}(x^{A}) &\defn - \int_{-\infty}^{\infty}~dV \int_{-\infty}^{V}dV^{\prime}E_{AB}(V^{\prime},x^{A}) \nonumber \\
&= \frac{1}{2}\lb[\gamma_{AB}\big\vert_{V=\infty} - \gamma_{AB}\big\vert_{V=-\infty}\rb]
\end{align}
where $E_{AB}=-\frac{1}{2}\partial_{V}^{2}\gamma_{AB}$ are the angular components of the (gauge-invariant) electric Weyl tensor and on the right-hand side we have chosen a gauge where $\gamma_{Vb}\vert_{\mc{H}} = 0 =q^{AB}\gamma_{AB}\vert_{\mc{H}}$ where $q^{AB}$ in the inverse $2$-sphere metric on the horizon cross-sections. The horizon memory is gauge invariant and, if non-vanishing, implies that the field cannot decay to zero at both early and late times. The cosmological horizon memory\footnote{Additional arguments linking soft gravitons to the infrared behavior of quantum states in de Sitter have been given in \cite{Ferreira:2016hee,Ferreira:2017ogo,Mao:2024phf,Sloth:2025nan}.} is mathematically equivalent to similar memory effects that have been previously considered on black hole horizons \cite{Hawking:2016msc,Donnay:2018ckb,Rahman:2019bmk,Danielson:2022tdw} and at null infinity in asymptotically flat spacetimes \cite{Zeldovich:1974gvh,PhysRevLett.67.1486}.  

We first show that in the free theory, the vanishing of the memory yield a general gauge invariant criterion for the existence\footnote{The normalizability (or lack thereof) of $\Omega_{\BD}$ plays a critical role in recent studies of gravitationally dressed algebras and entropy in cosmology \cite{2023JHEP...02..082C,Witten:2023xze,Kudler-Flam:2024psh,Chen:2024rpx}.} of a de Sitter invariant vacuum $\Omega_{\BD}$. We show that the necessary and sufficient condition for the existence of $\Omega_{\BD}$ is that the memory commutes with all local, gauge invariant observables. If the local algebra does not satisfy this criterion then there exists a local observable which is conjugate to the memory. We show that the existence of this conjugate mode directly implies that $\Omega_{\BD}$ is not a normalizable state. Conversely, if the local algebra does satisfy this criterion we show, by explicit construction, that $\Omega_{\BD}$ exists. 

This general result reproduces the results of Allen and others for the free scalar field \cite{Allen:1985ux,PhysRevD.39.3642,Kirsten:1993ug,Bros:2010wa,Page:2012fn}. In sec.~\ref{sec:MMC}, for the massless, minimally coupled scalar we find that the algebra $\Alg$ of local observables commutes with all memories except the constant mode given by 
\begin{equation}
\Delta(x^{A}) = q
\end{equation}
where $q$ is a constant which is a conserved charge associated to the constant shift symmetry $\phi \to \phi +c$ of the massless scalar. We show that the subalgebra $\Alg_{\textrm{SI}}\subset \Alg$ of ``shift invariant'' observables commutes with the constant memory and so $\Omega_{\textrm{BD}}$ exists as a proper state on this subalgebra. However, the full algebra $\Alg$ contains field observables which do not commute with the memory. In particular, there exists\footnote{In fact, there exist an infinite number of conjugate observables labeled by the different solutions with constant memory (see appendix \ref{sec:covariant-stuff}).} a conjugate ``constant mode'' $p\in \Alg$ of $\phi$ satisfying
\begin{equation}
\label{eq:pqi}
[q,p]=i
\end{equation}
The de Sitter invariant vacuum has zero memory and has a uniform probability distribution over the conjugate constant mode $p$ \cite{Allen:1987tz,Kirsten:1993ug}. Thus, $\Omega_{\BD}$ is not a normalizable state on the full algebra \(\Alg\). In sec.~\ref{sec:HilbmmKG} we construct the physical Hilbert space $\Hilb_{\dS}$ of states on the full algebra \(\Alg\) which have a square-integrable ``wavefunction'' in the memory $q$ and its conjugate $p$ \cite{PhysRevD.39.3642,Kirsten:1993ug,Tolley:2001gg}. The Hilbert space \(\Hilb_\dS\) does carry a (strongly continuous) unitary representation of the de Sitter symmetry group, however there is no normalizable de Sitter invariant state in \(\Hilb_\dS\).\\

For the massive scalar field, electromagnetic field and the linearized gravitational field this above criterion yields a new, simple proof of the existence of $\Omega_{\BD}$. As opposed to previous approaches --- see, e.g., \cite{Allen:1985ux,Allen:1985wd,Allen:1986dd,Kleppe:1993fz,Faizal:2011iv,Park:2008ki,Floratos:1987ek,Higuchi:2000ge,Miao:2011fc,Mora:2012zi,Gerard:2024eef} --- our analysis is entirely gauge invariant and also does not rely on any analytical continuation or  Euclidean methods. In the case of the construction of the graviton vacuum, these latter methods have been previously called into question \cite{Miao:2010vs,Miao:2009hb,Gerard:2024eef}. Furthermore, the method of proof presented in this paper is essentially identical between the three different fields. In this sense, the existence of a de Sitter invariant graviton vacuum is on the same footing as the existence of the massive scalar and photon vacua which have long been known to admit a de Sitter invariant state \cite{Allen:1985ux,Allen:1985wd}. 

In sec.~\ref{subsec:mscalar} we consider the massive scalar field. The exponential decay of solutions to the massive Klein-Gordon equation in de Sitter with initial data of compact support in spacetime \cite{Dappiaggi:2008dk} implies that the right hand side of \eqref{eq:memscalarintro} vanishes. We show that this directly implies that 
\begin{equation}
\label{eq:Deltaphi}
[\Delta,\phi]=0
\end{equation}
where $\Delta$ is the memory and $\phi$ is the full, massive Klein-Gordon field. In otherwords, in contrast to \eqref{eq:pqi}, all ``modes'' of $\phi$ commute with the memory. Thus, as we show in sec.~\ref{subsec:mscalar}, it directly follows that one can quantize this space of solutions to obtain a Fock space with vacuum $\Omega_{\BD}$. 

In the electromagnetic case, we show that the decay of all solutions with initial data of compact support and the corresponding existence of $\Omega_{\BD}$ directly follows from gauge invariance of the algebra $\Alg_{\EM}$ of local observables. On any horizon, the memory generates large gauge transformations 
\begin{equation}
A_{A} \to A_{A} + \ms{D}_{A}\lambda(x^{A})
\end{equation}
where $\ms{D}_{A}$ is the covariant derivative on $\bb{S}^{2}$. Since all local gauge invariant observables are invariant under large gauge transformations it directly follows that 
\begin{equation}
\label{eq:gravmemintro}
[\Delta_{A},a]=0
\end{equation}
any $a\in \Alg_{\EM}$. We show that this directly implies that there exists a zero memory Fock space of states that decay at early and late times with vacuum $\Omega_{\BD}$. Integrating Maxwell's equations on the horizon yields
\begin{equation}
\label{eq:EMconst}
\ms{D}^{A}\Delta_{A} = E_r\bigg\vert_{V=-\infty}^{V=+\infty}+ \int_{-\infty}^{\infty} dV~ j_{V}.
\end{equation}
The first term in \eqref{eq:EMconst} is the change in the ``radial component'' of the electric field on the horizon (see sec.~\ref{subsec:EM} for the precise definition) and $j_{V}$ is any charge current flux through the horizon. In the free theory $j_{V}=0$m so the fact that the memory vanishes implies that $E_{r}$ must also decay at early and late times. Indeed, we independently show that $E_{r}$ exponentially decays at early and late times.  

The arguments in the gravitational case are identical to the arguments outlined above. The memory generates the large gauge transformation 
\begin{equation}
\gamma_{AB}\to \gamma_{AB}+ (\ms{D}_{A}\ms{D}_{B}-\frac{1}{2}q_{AB}\ms{D}^{2})\lambda(x^{A})
\end{equation}
on the horizon and so on the algebra $\Alg_{\GR}$ of local, gauge invariant observables satisfies eq.~\ref{eq:gravmemintro} with $\Delta_{A}$ replaced by $\Delta_{AB}$ and $a\in \Alg_{\GR}$. Similarly, integrating the linearized Einstein equation on the cosmological horizon yields 
\begin{equation}
\label{eq:memconstGR}
-\ms{D}^{A}\ms{D}^{B}\Delta_{AB} =  E_{rr}\bigg\vert_{V=-\infty}^{V=+\infty} + 8 \pi G_{\rm{N}} \int_{-\infty}^{\infty} dV~ T_{VV}
\end{equation}
where the first term is the change in the radial component of the perturbed Weyl tensor (see eq.~\ref{eq:Bianchisourcefree}) and $T_{VV}$ is the stress energy flux of any matter (or gravitational waves) through the horizon. In the free theory, we have that $T_{VV}=0$. This implies the existence of a (zero memory) Fock space of states with vacuum $\Omega_{\BD}$. Indeed, we explicitly construct this state.

{The above results establish that all quantum states of a free massive scalar, photon and graviton field must decay at asymptotically early and late times. This lack of infrared radiation implies that the existence of a normalizable Bunch-Davies vacuum and all radiation states lie in the standard (zero memory) Fock space $\Fock_{0}$. In sec.~\ref{sec:sources}, we begin to analyze the impact of sources on the infrared radiation. We work in the framework of a quantum theory coupled to a classical source. This serves as a controlled toy model for the fully interacting quantum theory.}

{In the presence of sources, we find that a large number of infrared particles can be produced. First, we note that Maxwell's equations on the horizon eq.~\ref{eq:EMconst} imply that when there is a current flux through a horizon, there is memory produced on that horizon unless the change in the radial component of the electric field cancels its contribution. Assuming that the source does not persist forever, the radial component decays and cannot cancel the contribution from the source. This failure of decay of the vector potential is an ``infrared tail''. Nevertheless, such a source will not produce an infinite number of soft photons. To see this, we consider a Cauchy slice to the future of the source and compute the (finite) number of photons in the state. This number is independent of the Cauchy surface on which we evaluate it. Because only a finite number of particles are produced, these radiation states lie in the standard Fock space. This is consistent with analyses with quantum sources \cite{Gorbenko:2019rza, Hollands:2010pr}.}

{Even though the source does not produce an infinite number of photons, we demonstrate that it can produce a large number of soft particles if it persists for times significantly longer than the Hubble scale. When this time is taken infinitely far to the future, we determine that an infinite number of particles can be produced and the radiation state lies in a unitarily inequivalent Fock space. We stress that this does not imply that the theory, when quantizing the sources, will have inequivalent representations. Indeed, we argue that this does not occur.}

The rest of the paper is organized as follows. In section~\ref{sec:dS}, we review the structure of de Sitter spacetime. In section~\ref{sec:MMC}, we provide the general quantization of a massless scalar field in de Sitter spacetime, illustrate the obstruction to the existence of the Bunch-Davies vacuum and construct the Hilbert space. In section \ref{sec:quantfreefields}, we generalize this criterion to general, free quantum fields by introducing the notion of horizon memory and prove that the Bunch-Davies vacuum exists for massive scalars, electromagnetic fields and linearized gravitational fields. In section~\ref{sec:sources}, we analyze the production of infrared radiation in the presence of sources.

We will generally follow the notation and conventions of \cite{Wald:1984rg}. We use abstract spacetime indices $a,b,c,\ldots,$ for tensors in the de Sitter spacetime. For tensors on the cosmological horizon $\mc{H}$ which are orthogonal to the null normal $n^{a}$, their components will be denoted with capital Latin indices $(A,B,C,\dots)$. For example, the pull-back of the electric field $E_{a}=F_{ab}n^{b}$ to $\mc{H}$ will be denoted as $E_{A}$. In the main text, quantum observables will be denoted by the boldfaced version of the symbol for the corresponding classical observable, e.g., for a classical scalar field $\phi$ the corresponding quantum field is denoted by $\op{\phi}$.

\section{de Sitter Spacetime}
\label{sec:dS}

In this section, we review the structure of de Sitter spacetime, its isometries, and useful coordinate systems. While our arguments generally apply to any $d$-dimensional de Sitter spacetime we will restrict, for definiteness, to $d=4$ dimensions. Four dimensional de Sitter spacetime can be viewed as a hyperboloid of spacelike position vectors \(X^\mu\) satisfying $X^\mu X_{\mu} =L^2$, in $5$-dimensional Minkowski spacetime, where $L$ is the de Sitter radius. This embedded hyperboloid is invariant under the Lorentz group of the {5}-dimensional  Minkowski spacetime and makes manifest the full $SO(4,1)$ isometry group of de Sitter spacetime.

It is convenient to introduce so-called \textit{global coordinates} \((t,x^i)\) which cover the entire de Sitter manifold 
\be
    X_0 = L\sinh(t/L) \eqsp X_i = L\cosh(t/L)x_i \eqsp x_ix^i = 1.
\ee
The constant time \(t\) hypersurfaces are $3$-spheres with coordinates \(x^i\). The induced line element on the de Sitter spacetime is
\be\label{eq:dS-metric}
    ds^2 = -dt^2 + L^2 \cosh^2(t/L)dS_{3}^2.
\ee
where \(dS_3^2\) is the line element on the unit 3-sphere \(\bb S^3\). The subgroup of the de Sitter group \(SO(4,1)\) which preserves the foliation given by \(t = \text{constant}\) surfaces is \(O(4)\).

\begin{figure}
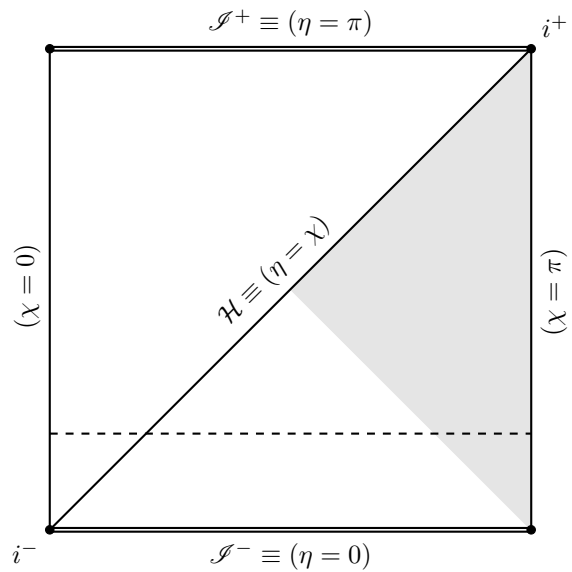

    \centering
    \include{penrose_ds_tikz}
    \caption{The Carter-Penrose diagram for global de Sitter spacetime. Each point on the diagram represents a $2$-sphere in spacetime, except on the vertical lines where it represents a point in the spacetime. In global coordinates (detailed in the text), a constant time slice is depicted by the horizontal dashed line representing a $3$-sphere. The ``doubled'' horizontal lines at the bottom and top of the diagram at $\eta = 0$ and $\eta = \pi$, represent past ($\scri^-$) and future ($\scri^+)$ conformal infinities, respectively. A cosmological horizon \(\hor\) corresponding to \(\eta = \chi\) is shown as a diagonal line. The gray shaded region is the associated static patch.}
    \label{fig:penrose_ds}
\end{figure}

It will be useful to define the \emph{conformal time} \(\eta \in (0,\pi)\) by
\be\label{eq:conf-time-defn}
    \tan \frac{\eta}{2} =  e^{{t}/{L}},
\ee
such that
\be\label{eq:dS-metric-conf}
    ds^2 = L^2 \csc^2 \eta \left(-d\eta^2 +dS_{3}^2\right).
\ee
This form shows that de Sitter spacetime is conformal to the cylinder \([0,\pi] \times \bb S^3\), i.e., a finite part of the Einstein static universe. The surfaces \(\eta = 0\) and \(\eta = \pi\) are the past and future conformal boundaries \(\scri^-\) and \(\scri^+\), respectively. In contrast to asymptotically-flat spacetimes these conformal boundaries are both spacelike. The corresponding Carter-Penrose conformal diagram is presented in \cref{fig:penrose_ds}.

For some of our main arguments we will also need to consider a cosmological horizon in the de Sitter spacetime. In global de Sitter spacetime there is no unique event horizon, but a cosmological horizon depends on the choice of some timelike geodesic or observer. Our arguments will apply to any choice of cosmological horizon and, for simplicity, we simply choose one cosmological horizon \(\hor\) as follows. Let us denote the polar coordinates on the unit 3-sphere cross-sections in the global foliation as \((\chi,\theta, \phi)\), where \(\chi, \theta \in [0,\pi]\) and \(\phi \in [0,2\pi)\). Our choice of cosmological horizon \(\hor\) is given by the null hypersurface defined by the equation \(\eta = \chi\). Note that \(\hor\) intersects the conformal boundaries \(\scri^-\) and \(\scri^+\) at the points \(i^- \equiv (\eta = \chi = 0)\) and \(i^+ \equiv (\eta = \chi = \pi)\), respectively. The action of the \(O(4)\) subgroup of the de Sitter group, which preserves the global foliation, maps the chosen horizon \(\hor\) to other possible choices for the cosmological horizon.

The affine parameter along the null generators of the horizon \(\hor\) is given by \(V = \tan(\eta - \pi/2)\) which runs from \(-\infty\) to \(+\infty\). This affine parameter \(V\) is the restriction to \(\hor\) of a null Kruskal coordinate in de Sitter spacetime; see, e.g., \cite{Hartong:2004rra}. The cross-sections of the horizon are diffeomorphic to the 2-sphere \(\bb S^2\), and we denote coordinates on these cross-sections by \(x^A\), e.g., \(x^A = (\theta,\phi)\) in polar coordinates.

\section{Massless Minimally Coupled Scalar Field}
\label{sec:MMC}
In this section, we consider the quantization of a massless, minimally coupled scalar field in de Sitter spacetime. In  sec.~\ref{subsubsec:IRdS} we isolate the ``infrared modes'' and the obstruction to the existence of a de Sitter invariant state. In sec.~\ref{subsubsec:shiftinvglob} we provide the Fock quantization of the ``shift invariant'' subalgebra in global coordinates which does admit a Bunch-Davies state. In sec.~\ref{sec:HilbmmKG} we provide the Hilbert space representation for the full scalar field.  

The standard approach to quantizing fields in de Sitter spacetime is to consider a Fock representation $\Fock_{0}$ with a de Sitter invariant vacuum state $\Omega_{\textrm{BD}}$. 
The field operator in this representation is decomposed in terms of a complete set of modes which suitably span $\Fock_{0}$. However, the basic issue we address is the quantization of fields in the possible absence of a (normalizable) ``ground state'' $\Omega_{\textrm{BD}}$. Therefore, it is extremely useful to provide a covariant quantization of fields on de Sitter spacetime which does not require an a priori choice of preferred state or Hilbert space. Such a formulation is provided by the algebraic approach to quantum field theory (see, e.g., \cite{Wald_1995,Hollands:2014eia,Witten:2021jzq} for further details). 

In any globally hyperbolic spacetime $(M,g)$, the collection of field observables we wish to quantize form a $\ast$-algebra $\Alg$. In addition to the identity $\op{1}$, the basic observable for the massless scalar field is the Klein-Gordon field $\op{\phi}$ which, due to vacuum fluctuations, is only defined in the distributional sense. Therefore, the basic element of $\Alg$ is the ``average'' of $\op{\phi}$ smeared with any real test function $f$ of compact support
\begin{equation}
\label{eq:phif}
\op{\phi}(f) = \int_{M}\sqrt{-g}d^{4}y~\op{\phi}(y)f(y)
\end{equation}
where $\sqrt{-g}d^{4}y$ is the measure on spacetime and $y$ are arbitrary coordinates on $M$. The observable $\op{\phi}$ is Hermition (i.e. $\op{\phi}^{\ast}=\op{\phi}$) and satisfies the massless, minimally coupled Klein-Gordon equation 
\begin{equation}
\label{eq:KGphi}
\Box_{g}\op{\phi}(y)=0
\end{equation}
where $\Box_{g}=g^{ab}\nabla_{a}\nabla_{b}$ and, here and below, all ``unsmeared'' relations are meant in the distributional sense. Finally, \eqref{eq:KGphi} is well-posed in de Sitter spacetime and so one can define unique retarded/advanced Green's functions. The covariant commutation relations are then given by 
\begin{equation}
\label{eq:commKG}
[\op{\phi}(y),\op{\phi}(y^{\prime})] = i E(y,y^{\prime})\op{1}
\end{equation}
where $E$ is the advanced-minus-retarded Green's function. The algebra $\Alg$ is the free algebra of smeared fields factored by the above distributional relations. 

The above covariant quantization can be directly related to the (more conventional) quantization of solutions to \eqref{eq:KGphi} on any Cauchy slice $\Sigma$. In particular, in de Sitter spacetime, $\Sigma$ could be a $t=~$constant surface in global coordinates or $\mc{H}^{+/-}$ the cosmological future/past horizon\footnote{The cosmological horizon does not admit the strict requirements of a Cauchy surface since a causal curve can ``miss'' the horizon by going to $i^{+/-}$. However, it is a good initial data surface since any smooth solution will not ``concentrate'' on these trajectories and will decay. Such solutions will thereby
have no symplectic flux through $i^{+}$.}. To see this, we note that the solutions to the Klein-Gordon equation are endowed with a symplectic form
\begin{equation}
\label{eq:sympprod}
\Omega_{\Sigma}(\phi_{1},\phi_{2})=\int_{\Sigma}~\sqrt{h}d^{3}x~n^{a}[\phi_{1}\nabla_{a}\phi_{2}-\phi_{2}\nabla_{a}\phi_{1}]
\end{equation}
where $\Sigma$ is a Cauchy surface and $n^{a}$ is the unit normal to the Cauchy surface. The symplectic product is conserved, i.e., \eqref{eq:sympprod} is independent of the Cauchy surface $\Sigma$. By lemma 3.2.1 of \cite{Wald_1995} , any field observable $\op{\phi}(f)$ can be expressed in terms of its corresponding ``initial data'' by 
\begin{equation}
\label{eq:phisympprod}
\op{\phi}(f) = \Omega_{\Sigma}(\op{\phi},Ef)
\end{equation}
where $Ef$ is the advanced-minus-retarded solution of \eqref{eq:KGphi} with source $f$ 
\begin{equation}
Ef(y) \defn \int_{M}\sqrt{-g}d^{4}y^{\prime}~E(y,y^{\prime})f(y^{\prime}).
\end{equation}
The commutation relations on $\Sigma$ can be expressed as 
\begin{equation}
\label{eq:commKGSigma}
[\op{\phi}(f_{1}),\op{\phi}(f_{2})] = i\Omega_{\Sigma}(Ef_{1},Ef_{2})\op{1}.
\end{equation}
Using \eqref{eq:sympprod} and \eqref{eq:phisympprod} one can obtain the more conventional decomposition of $\op{\phi}$ on $\Sigma$ 
\begin{equation}
\op{\phi}(f) = \op{\varphi}(n^{a}\nabla_{a}Ef) + \op{\pi}(Ef)
\end{equation}
where, formally, $\op{\varphi}=\op{\phi}\vert_{\Sigma}$ and $\op{\pi}=[n^{a}\nabla_{a}\op{\phi}]\vert_{\Sigma}$ which, using \ref{eq:commKGSigma}, satisfies the canonical commutation relations $[\op{\varphi}(x_{1}),\op{\pi}(x_{2})] = i \delta_{\Sigma}(x_{1},x_{2})\op{1}$ where $\delta_{\Sigma}$ is the delta function on $\Sigma$. 

While the above quantization defines the algebra of observables, to do physics, we not only need an algebra but also states. A state $\Psi$ is simply a positive, normalized linear function on $\Alg$, i.e., $\Psi:\Alg \to \bb{C}$ such that $\Psi(\op a^{\ast} \op a)\geq 0$ for all $\op a\in \Alg$ and $\Psi(\op{1})=1$. Since the algebra is generated by $\op{\phi}(f)$ a state can equivalently be specified by smearing its $n$-point correlation functions $\Psi\lb(\op{\phi}(y_{1})\dots \op{\phi}(y_{n})\rb)$. An important class of states is known as ``Gaussian states'' (or ``quasi-free states'' or ``vacuum states'') whose $n$-point functions are uniquely determined by their $1$-point functions $\Psi\lb(\op{\phi}(y)\rb)$ and $2$-point function $\Psi\lb(\op{\phi}(y_{1})\op{\phi}(y_{2})\rb)$, (see \cite{Hollands:2014eia}). 

The notion of an algebraic state is equivalent to the more conventional definition of a state as a vector in a Hilbert space by the \emph{GNS construction} \cite{GNS1,GNS2,Hollands:2014eia}. 
Indeed, given any algebraic state $\Psi$, one can define an inner product 
on the algebra $\braket{\op a| \op b}\defn \Psi(\op a^* \op b)$ for any $\op a, \op b\in \Alg$ and factoring by any degenerate elements. Completing the algebra with respect to this inner product yields a Hilbert space $\Hilb$ with $\ket{\Psi}\defn \ket{\op{1}}\in \Hilb$ in this representation. While the definition of an algebraic state and a Hilbert space state are equivalent, the notion of an algebraic state allows one to define the quantum theory without specifying a preferred state or Hilbert space representation a priori. If $\Psi$ is a Gaussian state, then the corresponding Hilbert space representation is a Fock space $\Fock$ \cite{Kay_1988}. Additionally, one also must impose a physical criteria the ``ultraviolet behavior'' of the correlation functions $\braket{\Psi | \op{\phi}(y_{1})\dots \op{\phi}(y_{n}) | \Psi}$. Roughly speaking, the \emph{Hadamard condition} requires that all physical states ``look like'' the Minkowski vacuum at sufficiently short distances. We refer the reader to, e.g., \cite{Radzikowski:1996pa,Hollands:2014eia}, for the precise, covariant definition of this condition. In this paper, we will primarily address infrared issues in the specification of a state and all states considered will be Hadamard.
\subsection{Quantization in de Sitter: infrared modes and the ``shift invariant algebra''}
\label{subsubsec:IRdS}

While the high frequency behavior of any state is universal, the infrared behavior is controlled by the low frequency behavior of the solutions to the Klein-Gordon equation and strongly depends upon the spacetime under consideration. In particular, in de Sitter spacetime, two features of the solutions of the massless, minimally coupled Klein-Gordon equation play a prominent role in this paper. The first is that, in de Sitter spacetime, the massless minimally coupled Klein-Gordon equation admits a constant solution $\phi=c$. The second key property is that if $\phi$ is any classical solution then the quantity 
\begin{equation}
\label{eq:charge}
q[\phi] \defn \int_{\Sigma}\sqrt{h}d^{3}x~n^{a}\nabla_{a}\phi 
\end{equation}
is a conserved ``charge'', i.e., \cref{eq:charge} is independent of the chosen Cauchy surface $\Sigma$. In the quantum theory, the  charge observable $\op{q}$ is defined as the observable where the scalar field is symplectically-smeared with the unit constant solution 
\begin{equation}
\label{eq:chargeops}
\op{q}\defn -\Omega_{\Sigma}(\op{\phi},1) = \int_{\Sigma}\sqrt{h}d^{3}x~n^{a}\nabla_{a}\op{\phi}.
\end{equation}
Using \eqref{eq:phisympprod} and \eqref{eq:chargeops}, it is straightforward to show that the local observables $\op{\phi}(f)$ have a non-vanishing commutation relation with the charge $\op{q}$ 
\begin{equation}
\label{eq:qcommphi}
[\op{q},\op{\phi}(f)] = i\Omega_{\Sigma}(1,Ef)\op{1} = iq[Ef]\op{1}
\end{equation}
which is non-vanishing if and only if the charge $q[Ef]\neq 0$. To see that this charge is generally non-vanishing, we first consider the retarded solution $E^{+}f$ which satisfies 
\begin{equation}
\label{eq:retf}
\Box_{g}(E^{+}f) = f.
\end{equation}
Let $D$ be the subregion of de Sitter spacetime that is topologically $[0,1]\times \bb{S}^{3}$, bounded by the spacelike Cauchy surfaces $\Sigma_{0}$ and $\Sigma_{1}$ where $\Sigma_{0}$ is to the past of $\Sigma_{1}$ and the support of $f$ does not intersect either $\Sigma_{0}$ or $\Sigma_{1}$. Integrating both sides of \eqref{eq:retf} over $D$ yields 
\begin{align}
\label{eq:retf12}
&\int_{\Sigma_{1}}\sqrt{h}d^{3}x~n^{a}\nabla_{a}(E^{+}f) - \int_{\Sigma_{0}}\sqrt{h}d^{3}x~n^{a}\nabla_{a}(E^{+}f)  \nonumber \\
&=\int_{D}\sqrt{-g}d^{4}y~f(y)
\end{align}
The first term is the charge of $E^{+}f$ on $\Sigma_{1}$, the second term vanishes since $E^{+}f$ vanishes in past of the support of $f$ and the right hand side of \eqref{eq:retf12} converges since $f$ is of compact support. It therefore follows that 
\begin{equation}
\int_{\Sigma}\sqrt{h}d^{3}x~n^{a}\nabla_{a}(E^{+} \bar{f}) = \int_{D}\sqrt{-g}d^{4}y~f(y)
\end{equation}
for any $\Sigma$ to the future of $\textrm{supp}(f)$. An identical argument shows that the advanced solution $(E^{-} {f})$ has charge 
\begin{equation}
\int_{\Sigma}\sqrt{h}d^{3}x~n^{a}\nabla_{a}(E^{-} {f}) = -\int_{D}\sqrt{-g}d^{4}y~f(y)
\end{equation}
on any $\Sigma$ to the past of $\textrm{supp}(f)$. The advanced minus retarded solution is a source-free solution of the Klein-Gordon equation and it follows that 
\begin{equation}
\label{eq:chargeEf}
q[Ef] = -\int_{M}\sqrt{-g}d^{4}y~f(y)
\end{equation}
which is generally non-vanishing for a function $f$ of compact support. 

Before proceeding towards the detailed arguments of this paper it is straightforward to foresee a contradiction between (1) the existence of a de Sitter invariant Bunch-Davies vacuum $\Omega_{\BD}$ and (2) that $\Omega_{\BD}$ is a state on $\Alg$. As we will show in the following subsection, $\Omega_{\BD}$ is annihilated by $\op{q}$ and therefore, in addition to being de Sitter invariant, it is also shift-invariant\footnote{In sec.~\ref{subsubsec:shiftinvglob} we construct the eigenstates of $\op{q}$ and show that $\Omega_{\BD}$ is an eigenstate with vanishing eigenvalue. This property has also been shown by considering the $m\to 0$ limit of $\Omega_{\BD}$ in the massive Klein-Gordon theory \cite{Garriga:2007zk}. Additionally, in sec.~\ref{subsec:MMCredux} we will show that states with non-vanishing $q$ have non-vanishing ``horizon memory'' (i.e. a change in the value of the field at early and late affine times) and thereby explicitly break de Sitter invariance.}. This fact has also been noted previously by several authors, e.g., \cite{Allen:1987tz,Kirsten:1993ug,Page:2012fn}.
Thus we have that
\begin{equation}
\braket{\Omega_{\BD} | \op{q}\op{\phi}(f)| \Omega_{\BD}} = \braket{\Omega_{\BD} | \op{\phi}(f)\op{q} |\Omega_{\BD}} = 0 
\end{equation}
and so
\begin{equation}
0=\braket{\Omega_{\BD}|[\op{q},\op{\phi}(f)]|\Omega_{\BD}} = iq[Ef] \braket{\Omega_{\BD}|\op{1}|\Omega_{\BD}}
\end{equation}
If the Bunch-Davies state is normalizable then the right-hand side is equivalent to $q[Ef]$ which, as we have just shown, is non-vanishing for general $f$. Thus, the Bunch-Davies state cannot be normalizable on the full algebra.\footnote{{We note that this non-normalizability crucially depended on the non-compact target space of the field. For example, if we instead took the scalar field to be circle-valued, the charge would be quantized and the Bunch-Davies state would be normalizable.}}

The fundamental issue in the above argument is the assumption that the Bunch-Davies state is a normalizable state on $\Alg$. This general argument that $\Omega_{\BD}$ is non-normalizable is in complete agreement with the result of Allen that there is no de Sitter invariant state on $\Alg$ \cite{Allen:1985ux}. Indeed, in sec.~\ref{sec:HilbmmKG}, we will explicitly construct the Hilbert space of states $\Hilb_{\dS}$ which is an irreducible representation of $\Alg$ and contains $\Omega_{\BD}$ as a non-normalizable, improper state.

 However, the Bunch-Davies vacuum can be defined on the subalgebra of observables with vanishing charge $q$ \cite{Bros:2010wa}. This algebra can be isolated by restricting the class of test functions to functions that satisfy
\begin{equation}
\label{eq:divf}
\bar{f}\defn \nabla_{a}f^{a}
\end{equation}
where $f^{a}$ is a test vector field of compact support. The field observable can be expressed as 
\begin{equation}
\op{\phi}(\bar{f}) = \int_{M}\sqrt{-g}d^{4}y~\op{\phi}(y)\bar{f}(y) = -\int_{M}\sqrt{-g}d^{4}y~\nabla_{a}\op{\phi}(y)f^{a}(y)
\end{equation}
Quantizing this subalgebra is equivalent to quantizing $\nabla_{a}\op{\phi}$ which is invariant under the constant shift $\op{\phi} \to \op{\phi} + c$. We will generally denote the sublagebra generated by $\op{\phi}(\bar{f})$ as $\Alg_{\NCM}$ where the subscript refers to ``shift invariant.'' By \eqref{eq:chargeEf}, the charge $q[E\bar{f}]$ vanishes and therefore
\begin{equation}
[\op{q},\op{\phi}(\bar{f})] = 0 
\end{equation}
and the above obstruction to the existence of $\Omega_{\BD}$ does not apply to $\Alg_{\NCM}$, in agreement with the result of Page and Wu \cite{Page:2012fn}. Finally, we may trivially extend the algebra to $\Alg_{q,\NCM}$ to include the charge observable. We will generally refer to $\Alg_{q,\NCM}$ as the ``shift-invariant'' algebra.

\subsection{Fock quantization of the shift-invariant algebra: global slicing}

\label{subsubsec:shiftinvglob}
We first present the explicit quantization of $\Alg_{q,\NCM}$ in global coordinates on any constant $t$ hypersurface $\Sigma \cong \bb{S}^{3}$. The symplectic form \eqref{eq:sympprod} on this hypersurface is given by 
\be
    \Omega_{\Sigma}(\phi_1,\phi_2) = \int_{\bb S^3} d\Omega_3 ~ \cosh^3 t \lb( \phi_1 \partial_t \phi_2 - \phi_2 \partial_t \phi_1 \rb)
\ee
In these coordinates, it is convenient to note that one can decompose any solution $\phi(y)$ into a complete set of modes  \cite{Allen:1987tz}
\begin{equation}
\label{eq:modeglobal}
\phi(y) = c_{1} + c_{2}\xi(t) + \sum_{k\neq 0}c_{k\ell m}u_{k\ell m}(y)
\end{equation}
where $c_{1},c_{2},c_{k\ell m}\in \bb{R}$, $\xi(t)$ is the spherically symmetric solution
\begin{equation}\label{eq:xi-defn}
\xi(t)= \frac{1}{2\pi^{2}}\int_0^t \frac{dt'}{\cosh^3 t'} = \frac{1}{4\pi^{2}} \lb[ \frac{\sinh t}{\cosh^2 t} + \tan^{-1}(\sinh t) \rb]. 
\end{equation}
and  $u_{k\ell m}(y) = u_{k}(t)Y_{k\ell m}(x^{i})$ where $Y_{k\ell m}$ are the spherical harmonics on $\bb{S}^{3}$. The functions $u_{k}(t)$ can be simply expressed in terms of the conformal time $\eta$ (see \cref{eq:conf-time-defn}) as \cite{Allen:1987tz}
\begin{equation}
\label{eq:modezeromass}
u_{k}(\eta)=\sin^{\nfrac{3}{2}} \eta \lb[ A_k P_{k + \half}^{\nfrac{3}{2}}(-\cos \eta) + B_k Q_{k + \half}^{\nfrac{3}{2}}(-\cos \eta) \rb] 
\end{equation}
where $P_{k+\half}^{\nfrac{3}{2}}(x)$ and $Q_{k+\half}^{\nfrac{3}{2}}(x)$ are associated Legendre functions of the first and second kind defined in the interval \(x \in (-1,1)\). The modes \(u_{k\ell m}\) are ortho-normalized in the (conserved) Klein-Gordon norm \(\inp{\phi_1,\phi_2}_\KG \defn -i \Omega_\Sigma(\phi_1,\phi_2^*) \) so that 
\begin{equation}
\inp{u_{k\ell m}, u_{k'\ell' m'}}_\KG = \delta_{k,k'}\delta_{\ell\ell'}\delta_{m,m'}
\end{equation}
and $\inp{u_{k\ell m}, u^*_{k'\ell' m'}}_\KG = 0$. This fixes the constants \(A_k\) and \(B_k\) to be
\be
    A_k = \frac{\pi}{2}i B_k = e^{i 3\pi /4} \lb( \frac{\pi}{4} \frac{\Gamma(k)}{\Gamma(k + 3)}\rb)^\half.
\ee

In this decomposition, the solutions with non-vanishing charge can be straightforwardly parameterized. Indeed, the constant mode $c_{1}$ and the modes $u_{k\ell m}(y)$ all have vanishing charge 
\begin{equation}\label{eq:charge-u}
q[u_{k\ell m}] = \int_{\bb{S}^{3}} d \Omega_{3} \cosh^{3}t~\partial_{t}u_{k\ell m}  = 0.  
\end{equation}
for any $k,\ell,m$ with $k>0$. However, the mode $\xi$ has unit charge 
\begin{equation}
\label{eq:chargexi}
q[\xi]=\int_{\Sigma}\sqrt{h}d^{3}x~n^{a}\nabla_{a}\xi(t) = \int_{\bb{S}^{3}} d \Omega_{3} \cosh^{3}t~\partial_{t}\xi(t) = 1
\end{equation}
where we used the fact that $\partial_{t}\xi = (1/2\pi^{2})\sech^{3}t$.

We now consider the irreducible representations of the ``shift-invariant'' subalgebra $\Alg_{q,\NCM}$. Since $\op{q}$ commutes with all elements of $\Alg_{\NCM}$ it follows from Schur's lemma that this is equivalent to obtaining the irreducible representations of $\Alg_{\NCM}$ and, in each irreducible representation, $\op{q}=q\op{1}$ is a multiple of the identity. 
To obtain these irreducible representations we start with the Bunch-Davies vacuum state on $\Alg_{\NCM}$ which is the de Sitter invariant, Gaussian state with vanishing one-point function and two-point function given by 
\be
&\braket{\Omega_{\BD}| \op{\phi}(\bar{f}_{1})\op{\phi}(\bar{f}_{2}) |\Omega_{\BD}} \\
& \qquad = \int_{M^2}d^4y_1 d^4y_2~ \braket{\Omega_{\BD} | \op{\phi}(y_{1})\op{\phi}(y_{2}) |\Omega_{\BD}}\bar{f}_{1}(y)\bar{f}_{2}(y_{2})
\ee
for any $\bar{f}_{1},\bar{f}_{2}$. To give an explicit expression for the two-point function we can expand $\Omega_{\BD}$ in terms of modes given by \eqref{eq:modeglobal}. Since the solutions $E\bar{f}$ have vanishing charge and the test functions $\bar{f}=\nabla_{a}f^{a}$ project out the constant mode, it follows that the Bunch-Davies vacuum on $\Alg_{\NCM}$ can be expressed entirely in terms of the modes $u_{k\ell m}$. Indeed, it has been shown that the Gaussian state defined by the $2$-point function 
\begin{equation}
\braket{\Omega_{\BD}| \op{\phi}(y_{1})\op{\phi}(y_{2})|\Omega_\BD} = \sum_{k \neq 0}u_{k\ell m}(y_1)u^{\ast}_{k\ell m}(y_2)
\end{equation}
is positive, de Sitter invariant, Hadamard and satisfies the wave equation as well as the commutation relation when smeared with any test functions $\bar{f}$ \cite{2007rqft.book...27B,Bros:2010wa}. Thus $\Omega_{\BD}$ is a well-defined state on the subalgebra $\Alg_{\NCM}$. The GNS Fock space corresponding to this Gaussian state is denoted as $\Fock_{0}$. The subscript ``$0$'' denotes the fact that all states in $\Fock_{0}$ have zero charge 
\begin{equation}
\op{q}\ket{\Psi_{0}} = 0 \textrm{ for any $\ket{\Psi_{0}}\in \Fock_{0}$}.
\end{equation}
This directly follows from the fact that $\op{q}$ is a multiple of the identity on $\Fock_{0}$ and one can straightforwardly check that $\braket{\Psi | \op{q} | \Psi}$ vanishes for all $\ket{\Psi}\in \Fock_{0}$. This ``standard'' Fock representation admits a strongly continuous action of the de Sitter isometry group and can be decomposed in terms of a direct integral of unitary irreducible representations \cite{Mautner}. These representations have been completely classified in $d=2$ \cite{Pukánszky_1961,Repka_1978} and has recently been extended to $d=3$ with incomplete results for $d\geq 4$ \cite{Martin_1981,Penedones:2023uqc}.

One can straightforwardly generate an irreducible Fock representation $\Fock_{q}$ of $\Alg_{\NCM}$ consisting of states with charge $q$. Let $\Xi_{q}(y)$ be a smooth solution with charge $q$, {e.g.,~$q\xi(t)$} and, consider the automorphism $\aut_{\Xi_q}$ 
\begin{equation}
    \aut_{\Xi_q}: \op{\phi}(y) \mapsto \op{\phi}(y) + \Xi_q(y)\op{1} \,
\end{equation}
and $\aut_{\Xi_{q}}[\1] = \1$.
This map is straightforwardly extended to arbitrary sums of products of $\op{\phi}$. This automorphism defines a ``shifted'' vacuum \(\Omega_{\Xi_q}\) whose expectation value of any $\op{O}\in \Alg$ is 
\begin{equation}
\braket{\Omega_{\Xi_{q}}|\op{O} |\Omega_{\Xi_{q}}} \defn \braket{\Omega_{\BD}| \aut_{\Xi_{q}}[\op{O}]|\Omega_{\BD}}.
\end{equation}
Thus, the state $\Omega_{\Xi_{q}}$ corresponds to a vacuum with non-vanishing vacuum expectation value $\braket{\Omega_{\Xi_{q}}|\op\phi(y)|\Omega_{\Xi_{q}}} = \Xi_{q}(y) \neq 0$. If $\Xi_{q}(y)$ is a solution with vanishing charge $q=0$, then $\Omega_\Sigma(\op\phi, \Xi_q) \in \Alg_{\NCM}$, and the automorphism \(\aut_{\Xi_{q}}\) can be implemented on $\Fock_{0}$ by the unitary operator $e^{i \Omega_\Sigma(\op\phi, \Xi_q)}$. The algebraic state \(\Omega_{\Xi_q}\) is given by the state vector $\ket{\Omega_{\Xi_q}} = e^{i \Omega_\Sigma(\op\phi, \Xi_q)}\ket{\Omega_{\BD}}$ which is a coherent state in $\Fock_{0}$.

However, if $q\neq 0$ then $\Xi_q(y)$ cannot be expanded purely in terms of the mode functions $u_{k\ell m}$ and $\ket{\Omega_{\Xi_q}}$ is {\em not} a state in $\Fock_{0}$. The GNS construction with respect to $\ket{\Omega_{\Xi_q}}$ yields a Fock space $\Fock_{\Xi_q}$ which is unitarily inequivalent to $\Fock_{0}$. Furthermore, it is straightforward to check that if $\Xi_q(y)$ and $\Xi_q'(y)$ are both smooth solutions with the same charge $q$ then their corresponding GNS representations $\Fock_{\Xi_q}$ and $\Fock_{\Xi_q'}$ are unitarily equivalent with unitary map $\exp\big(i\Omega_\Sigma(\op{\phi}, \Xi_q-\Xi_q')\big)$. Therefore, up to unitary equivalence, each Fock space is labeled by the charge $q$ --- instead of the representative solution \(\Xi_q(y)\) --- and all states in $\Fock_{q}$ are eigenstates of the charge operator with eigenvalue $q$
\begin{equation}
\op{q}\ket{\Psi_{q}} = q\ket{\Psi_{q}}  \textrm{ for any $\ket{\Psi_{q}}\in \Fock_{q}$}.
\end{equation}
Thus, there are uncountably-many irreducible representations $\Fock_{q}$ of $\Alg_{\rm{SI},q}$ labeled by the charge $q$. Finally, we note that the observable $q$ is de Sitter invariant and consequently, the de Sitter group has a strongly, continuous action on each Fock space $\Fock_{q}$.

\subsection{The Hilbert space of a massless, minimally coupled scalar}
\label{sec:HilbmmKG}

We now consider the representations of the full algebra $\Alg$ which amounts to including solutions $Ef$ with non-vanishing charge. We note that the Fock spaces $\Fock_{q}$ of definite $q$ cannot be representations of the full algebra $\Alg$ since, by \eqref{eq:qcommphi}, $\op{\phi}(f)$ does not commute with $\op{q}$ if $q[Ef]\neq 0$. 

To obtain a representation of $\Alg$ it is useful to utilize the mode decomposition \eqref{eq:modeglobal} in the global coordinates since the charged solutions are conveniently parameterized by the mode $\xi(t)$ which has unit charge, i.e., $q[\xi]=1$. The covariant formulation of the following construction, without using any preferred foliation, is given in \cref{sec:covariant-stuff}. Thus, with respect to the decomposition of \eqref{eq:modeglobal}, quantizing the full algebra is equivalent to adding the quantum observable 
\begin{equation}
\label{eq:p-xi-defn}
\op{p}(\xi) \defn -\Omega_{\Sigma}(\op{\phi},\xi)
\end{equation}
to the algebra $\Alg_{q,\NCM}$. Since $\xi$ is symplectically orthogonal to the modes $u_{k\ell m}$, it follows that $\op{p}(\xi)$ commutes with the subalgebra $\Alg_{\NCM}$. However, since $\xi$ has unit charge, $\op{p}(\xi)$ does not commute with $\op{q}$. In summary, 
\begin{equation}
\label{eq:commpq}
[\op{p}(\xi),\op{\phi}(\bar{f})]=0,\quad \quad [\op{q},\op{p}(\xi)] = i\op{1}
\end{equation}
for any $\bar{f}$ satisfying \eqref{eq:divf}. 

Each Fock space $\Fock_{q}$ is an eigenspace of $\op{q}$ and so the conjugate $\op{p}(\xi)$ cannot be represented on any Fock representation $\Fock_q$. The situation is analogous to the case of one-dimensional quantum mechanics where the position operator has a well-defined action on position eigenstates, but a representation of the momentum operator can only be obtained if one suitably integrates over position eigenstates on the real line. Similarly, one can obtain a representation $\Hilb_{\dS}$ of the full algebra by ``integrating'' over the Fock spaces $\Fock_{q}$ with respect to $q$ with the Lebesgue measure\footnote{The analogue of the Stone-von Neumann theorem, applied to the operators \(\op q\) and \(\op p(\xi)\), shows that this Hilbert space is unique up to unitary equivalence.} on $\bb{R}$. In particular, we define a state $\Psi \in \Hilb_{\dS}$ by a specification of a (measurable) family of states $\psi(q)\in \Fock_{q}$ such that 
\be\label{eq:normdS}
    \norm{\Psi}^{2}\defn \int_{\bb{R}}dq~ \norm{\psi(q)}^{2}_{q}<\infty
\ee
where the integrand is the norm of each $\psi(q)$ obtained from the inner product $\braket{\cdot|\cdot}_{q}$ on $\Fock_{q}$. The inner product of two such states $\Psi_{1},\Psi_{2}\in \Hilb_{\dS}$ is given by 
\be
\braket{\Psi_{1}|\Psi_{2}}\defn \int_{\bb{R}}dq~\braket{\psi_{1}(q)|\psi_{2}(q)}_{q}.
\ee
Taking the completion of such states with respect to the norm \eqref{eq:normdS} yields the Hilbert space $\Hilb_{\dS}$ which is a separable ``direct integral'' Hilbert space denoted as 
\begin{equation}
\label{eq:Hilbds}
\Hilb_{\dS} = \int_{\bb{R}}^{\oplus}dq~\Fock_{q}.
\end{equation}
This direct integral is isomorphic to $\Fock_0 \otimes L^2(\mathbb{R})$ (see \cref{rem:DI-TP}).
This representation is an irreducible representation of the full field algebra $\Alg$. We note that $\Alg_{q,\NCM}$ has a well-defined action on each Fock space $\Fock_{q}$ and therefore this subalgebra is densely defined on $\Hilb_{\dS}$. In particular, given any state $\Psi \in \Hilb_{\dS}$ corresponding to a family of states $\ket{\psi(q)}\in \Fock_{q}$ for any $q$, the action of $\op{q}$ corresponds to a multiplication operator 
\begin{equation}
\op{q}\ket{\Psi}\defn q \ket{\psi(q)}
\end{equation}
and so is densely defined on the domain of states $\ket{\Psi}=\ket{\psi(q)}$ such that 
\begin{equation}
\int_{\bb{R}} dq~q^{2}||\psi(q)||^{2}_{q}<\infty. 
\end{equation}
Additionally, $\Hilb_{\dS}$ admits a strongly continuous, unitary action of the de Sitter group inherited from its action on each ``fibre'' $\Fock_{q}$ in the direct integral. However, there is no de Sitter invariant proper state in \(\Hilb_\dS\).

It remains to be shown that the  ``conjugate'' mode $\op{p}(\xi)$ is also densely defined on $\Hilb_{\dS}$. This mode does not commute with $\op{q}$ and so, in analogy with one-dimensional quantum mechanics, one expects that, on $\Hilb_{\dS}$, $\op{p}(\xi)$ acts as a ``derivative operator with respect to $q$''. However, unlike in quantum mechanics, the states are not complex-valued functions in $\bb{R}$ but are valued in the Fock spaces $\Fock_{q}$ at each $q$. Thus, we must exercise some care in defining $\op{p}(\xi)$. 

The Fock spaces $\Fock_{q}$ were constructed by shifting the vacuum expected value of $\op{\phi}$ in $\Omega_{\BD}$ by an arbitrary solution $\Xi_{q}(y)\op{1}$ with charge $q$ which yield a family of vacua $\Omega_{\Xi_{q}}$. To specify the action of $\op{p}(\xi)$ on $\Hilb_{\dS}$, it will be convenient to choose the family of vacua obtained by $\Xi_{q}=q\xi(t)$ which we denote as $\ket{\Omega_{q\xi}}$. This choice can be made without loss of generality since, as previously stated, any other choice of family of vacua can be related to this choice by a unitary. The general action of $\op{p}(\xi)$ for an arbitrary choice of vacuum can be found in appendix~\ref{sec:covariant-stuff}. Given this choice of sequence of vacua, we consider the following states in $\Hilb_{\dS}$ 
\be
\label{eq:Psidense}
\ket{\Psi} = \sum_{n=0}^{\infty}\int_{M^{n}}~d^{4}y_{1} &\dots d^{4}y_{n}~\psi(y_{1},\dots,y_{n};q) \\
&\times \op{\phi}(y_{1})\dots\op{\phi}(y_{n})\ket{\Omega_{q\xi}}
\ee
where $\psi(y_{1},\dots,y_{n};q)$ is smooth in all of its variables, has vanishing charge and, decays sufficiently rapidly such that $\ket{\Psi}$ has finite norm. States $\ket{\Psi}$ of the form of \eqref{eq:Psidense} are dense in $\Hilb_{\dS}$ and therefore it suffices to define the action of $\op{p}(\xi)$ on such states. We begin by defining the action of the unitary $\exp(is\op{p}(\xi))$ for any $s\in \bb{R}$. By eq.~\eqref{eq:commpq}, this unitary has a trivial action on $\op{\phi}(\bar{f})\in \Alg_{\NCM}$ and shifts the charge $\op{q}$ by $s\op{1}$
\begin{equation}
e^{-is\op{p}(\xi)}\op{q}e^{is\op{p}(\xi)} = \op{q}+ s\op{1}.
\end{equation}
Therefore, the action of this unitary on any $\ket{\Psi}$ of the form of \eqref{eq:Psidense} is 
\be
\label{eq:Psidenseexpp}
e^{is \op{p}(\xi)}\ket{\Psi} = \sum_{n=0}^{\infty}\int_{M^{n}}~d^{4}y_{1} &\dots d^{4}y_{n}~\psi(y_{1},\dots,y_{n};q-s)\times  \\
&\times \op{\phi}(y_{1})\dots\op{\phi}(y_{n})\ket{\Omega_{q\xi}}. 
\ee
The infinitesimal action of \(\op p(\xi)\) is obtained by differentiating \cref{eq:Psidenseexpp} at \(s = 0\)
\be 
&\op{p}(\xi)\ket{\Psi} \defn -i\frac{d}{ds}\bigg[e^{is \op{p}(\xi)}\ket{\Psi}\bigg]\bigg\vert_{s=0} \\ 
&= \sum_{n=0}^{\infty}\int_{M^{n}}~d^{4}y_{1}\dots d^{4}y_{n}\bigg[i \frac{\partial}{\partial q}\psi(y_{1},\dots,y_{n};q) \\
&\quad \quad \quad \quad \quad  \quad \quad \quad \quad \times \op{\phi}(y_{1})\dots\op{\phi}(y_{n})\ket{\Omega_{q\xi}}\bigg]
\ee 
This defines the action of $\op{p}(\xi)$ as a self-adjoint operator on a dense subspace of $\Hilb_{\dS}$.

As noted above, each Fock space \(\Fock_q\) has a strongly continuous action of the de Sitter group and thus, so does the direct integral \(\Hilb_\dS\). However, $\Hilb_{\dS}$ does not contain any de Sitter invariant proper state --- the Bunch-Davies state $\Omega_{\BD}$ is an improper state in this Hilbert space. This result is in complete agreement with the results of Allen \cite{Allen:1985ux} that there is no normalizable, de Sitter invariant state on $\Alg$. Nevertheless, we emphasize that the quantum field theory of the massless scalar field is well-defined and covariant under the de Sitter group. As an aside, a similar construction also holds for a massless scalar field in \(2\)-dimensional Minkowski spacetime considered in \cite{Ford:1985qh,Derezi_ski}.\footnote{If one instead considers a quantization which only has a well-defined action of the exponentiated Weyl operators $e^{i\op{q}}$ and $e^{i\op{p}}$ then $\Omega_{\BD}$ exists on the non-separable direct sum Hilbert space $\oplus_{q}\Fock_{q}$. The analogous construction for the $2$-dimensional massless scalar in Minkowski spacetime was given in \cite{Acerbi:1992yv,Acerbi:1993yu}.}

\begin{remark}[$\Hilb_\dS$ as a tensor product]\label{rem:DI-TP}
    The direct integral Hilbert space \(\Hilb_\dS\) can also be expressed (non-canonically) as the tensor product
    \begin{equation}
\Hilb_\dS \cong \Fock_0 \otimes L^2(\bb R).
    \end{equation}
 Let \(\xi(t)\) be the unit charge solution as in \cref{eq:xi-defn}, and \(f\) be an arbitrary test function. Then, according to \cref{eq:modeglobal,eq:charge-u} we can decompose the solution \(Ef\) as
    \be
        Ef = c_1 + c_2 \xi(t) + E\bar f
    \ee
    where $c_{1},c_{2}$ are constants and  \(E\bar f\) is a solution with vanishing charge. Using this decomposition we have 
    \be\label{eq:op-decomp}
        \op\phi(f) = \Omega_\Sigma(\op\phi, Ef) = -c_1 \op q - c_2 \op p(\xi) + \op\phi(\bar f) 
        %\eqsp \op\phi(f) \in \Alg, ~ \op\phi(\bar f) \in \Alg_\NCM
    \ee
    where $\op\phi(f) \in \Alg$ and $\op\phi(\bar f) \in \Alg_\NCM$. Now we identify the state \(\ket{\Omega_\BD} \otimes h(q) \in \Fock_0 \otimes L^2(\bb R)\) with the state in \(\Hilb_\dS\) given by the family of states \(h(q) \ket{\Omega_{q\xi}} \in \Fock_q\). Then using \cref{eq:op-decomp} the action of \(\op\phi(f) \in \Alg\) on \(\ket{\Omega_\BD} \otimes h(q)\) can be defined as
    \be
        &\op\phi(f) \bigg( \ket{\Omega_\BD} \otimes h(q) \bigg) \\
        &\quad\quad \defn \op\phi(\bar f) \ket{\Omega_\BD} \otimes \lb( - c_1 q  - i c_2 \frac{\partial}{\partial q} \rb) h(q) 
    \ee
    and this state can be identified with the state in \(\Hilb_\dS\) given by the family \(\op\phi(f) h(q) \ket{\Omega_{q \xi}} \in \Fock_q\). This is easily extended to the span of states obtained by multiple smeared operators \(\op\phi\) acting on \(\ket{\Omega_\BD} \otimes h(q)\) and it can be checked that this identification of \(\Fock_0 \otimes L^2(\bb R)\) with \(\Hilb_\dS\) is a unitary map. Note that this identification depends on the non-unique choice of a unit-charge solution $\xi(t)$ and thus is not unique.
\end{remark}

%%-----------------------------------------------------------------------------------
\section{Quantization of Free Fields}
\label{sec:quantfreefields}
In the previous section, we presented the quantization of a massless scalar field. We found that the algebra $\Alg$ does not admit a de Sitter invariant state. The basic obstruction to the existence of a de Sitter invariant state was the existence of a charge $\op{q}$ which annihilates $\Omega_{\BD}$ but does not commute the elements of $\Alg$. The purpose of this section is to establish a general criterion for the existence or non-existence of a de Sitter invariant vacuum state in more general free quantum field theories. We will illustrate that the charge $q$ is a special case of a more general observable on the horizon known as the ``horizon memory'' which controls the infrared behavior of any quantum state in de Sitter. The general obstruction to obtaining a de Sitter invariant state is the non-commutation of the algebra $\Alg$ of local fields with the horizon memory. 

We illustrate this by revisiting the quantization of the massless scalar on the horizon in sec.~\ref{subsec:MMCredux}. We then apply this criterion to a massive scalar field in sec.~\ref{subsec:mscalar}, the electromagnetic field in sec.~\ref{subsec:EM} and the linearized gravitational field in sec.~\ref{subsec:GR}. In particular we show that, in contrast to the massless scalar, the local observables have vanishing memory and these theories admit a de Sitter invariant vacuum.

\subsection{Quantization of the massless scalar on the cosmological horizon}
\label{subsec:MMCredux}

One can equivalently quantize the scalar field on the cosmological future horizon $\hor$. The initial data on $\hor$ of the Klein-Gordon equation corresponds to the specification of the field 
\begin{equation}
\Phi(V,x^{A}) \defn \lim_{\to \hor}\phi(y).
\end{equation}
The symplectic form \eqref{eq:sympprod} on $\hor$ is given by 
\begin{equation}
\label{eq:sympprodhor}
\Omega_{\hor}(\phi_{1},\phi_{2}) = \int_{\hor} dVd\Omega_{2}~(\Phi_{1}\partial_{V}\Phi_{2}-\Phi_{2}\partial_{V}\Phi_{1}).
\end{equation}
Letting 
\begin{equation}
F(V,x^{A}) = \lim_{\to \hor}Ef
\end{equation}
the algebra $\Alg$ restricted to $\hor$ is generated by the fields
\begin{equation}
\label{eq:Pi}
\op{\Pi}(F)\defn - \frac{1}{2} \Omega_{\hor}(\op{\phi},Ef) = - \frac{1}{2} \op{\phi}(f)
\end{equation}
for any test function $f$.  For any $F(V,x^{A})$ which vanishes as $V\to \pm \infty$, the observable can be expressed as
\begin{equation}
\op{\Pi}(F) = \int_{\hor}dVd\Omega_{2}~\partial_{V}\op{\Phi}(V,x^{A}) F(V,x^{A})
\end{equation}
 which is $\partial_{V}\op{\Phi}$ smeared on the horizon. Quantizing $\op{\phi}$ is equivalent quantizing the observables $\op{\Pi}$ subject to the commutation relations
\begin{equation}
\label{eq:Picomm}
[\op{\Pi}(F_{1}),\op{\Pi}(F_{2})]= - \frac{i}{4} \Omega_{\mc{H}}(F_{1},F_{2})\op{1}.
\end{equation}
The charge $q$ which played a prominent role in sec.~\ref{sec:MMC} controls the infrared behavior of the scalar field and on the horizon is given by 
\begin{equation}
q[\Phi] = \lim_{V\to \infty}\Phi(V,x^{A})\big\vert_{\ell =0} - \lim_{V\to -\infty}\Phi(V,x^{A})\big\vert_{\ell =0}
\end{equation}
where 
\begin{equation}
\Phi\vert_{\ell =0} \defn \frac{1}{4\pi}\int_{\bb{S}^{2}}d\Omega_{2}~\Phi(V,x^{A})
\end{equation}
is the spherically symmetric part of $\Phi$. More generally, the full infrared behavior of the field on the horizon is encoded in the ``horizon memory'' 
\begin{equation}
\label{eq:memphidiff}
\Delta(x^{A})=\lim_{V\to \infty}\Phi(V,x^{A}) - \lim_{V\to - \infty}\Phi(V,x^{A})
\end{equation}
where the spherically symmetric part satisfies $\Delta \vert_{\ell =0} = q$. We note that if $\Delta$ vanishes, then $\Phi$ at late times must decay to its early time value. Conversely, if $\Delta$ is non-vanishing, then this implies a permanent change in the value of $\Phi$ from early to late times on $\hor$. In contrast to $q$, the full horizon memory $\Delta$ is not conserved. However, the main arguments of this section are to show that the analogs of $\Delta$ vanish for the massive scalar, electromagnetic and linearized gravitational field algebras on any horizon. This property controls the global infrared behavior of quantum states in these theories. 

For the case of the massless scalar, it is straightforward to see that any source-free solutions $F=Ef\vert_{\hor}$ can only have, at most, constant memory, i.e., the memory $\Delta[F] = q[F]$ for any $F$. The modes $u_{k\ell m} \to - \sqrt{2/\pi}$ as $t \to \pm \infty$ and so, in addition to the constant mode, the memory of $u_{k\ell m}$ on any horizon vanishes 
\begin{equation}
\Delta[u_{k\ell m}] = 0.
\end{equation}
The only mode with non-vanishing memory is $\xi(t)$ which interpolates between $-1/8\pi$ and $+1/8\pi$ between early and late global times. Thus, on the horizon, the memory of $\xi$ is a constant 
\begin{equation}
\frac{1}{4\pi} \Delta[\xi]= q[\xi]= 1
\end{equation}
Thus while, for any $\lambda(x^{A})$, the full memory observable can be defined as
\begin{equation}
\label{eq:memscalar}
\Delta(\lambda) \defn -\Omega_{\hor}(\op{\phi},\lambda),
\end{equation}
the free massless scalar field has no higher harmonic memory and so we only need to explicitly quantize the charge $\op{q}$. However, quantizing the full memory observable $\op{\Delta}(\lambda)$ will be relevant in sec.~\ref{sec:sources} when we couple the field to a source. 

As before, the charge obeys the commutation relation
\begin{equation}
\label{eq:qPiF}
[\op{q},\op{\Pi}(F)] = - \frac{1}{2} [\op{q},\op{\phi}(f)] = - \frac{i}{2} q[F]\op{1}.
\end{equation}
The shift-invariant algebra $\Alg_{\NCM}$ is generated by the observables $\op{\Pi}(\bar{F})$ where $\bar{F}$ has vanishing memory and so $[\op{q},\op{\Pi}(\bar{F})]=0$. By the mode decomposition \eqref{eq:modeglobal} it follows that $\bar{F}$ on the horizon behaves as $\lim_{V\to \pm \infty}\bar{F} = \zeta(x^{A})$ where $\zeta$ is a smooth function on $\bb{S}^{2}$. The quantization of this space of initial data is equivalent to the quantization of initial data $\bar{F}_{\zeta} = \bar{F}-\zeta $ which vanish as $V\to \pm \infty$ since the observable $\op{\Pi}(\bar{F})$ differs from $\op{\Pi}(\bar{F}_{\zeta})$ by $\op{\Delta}(\zeta)$ which commutes with all elements of $\Alg_{\NCM}$. Thus, to quantize $\Alg_{\NCM}$, we may simply restrict to the initial data $\bar{F}$ on $\hor$ which satisfy 
\begin{equation}
\lim_{V\to \pm \infty}\bar{F}(V,x^{A}) = 0.
\end{equation} 
Using \cref{eq:sympprodhor,eq:Picomm}, the elements of $\Alg_{\NCM}$ satisfy
\begin{equation}
[\op{\Pi}(x_{1}),\op{\Pi}(x_{2})] = \frac{i}{2}\delta^{\prime}(V_{1},V_{2})\delta_{\bb{S}^{2}}(x_{1}^{A},x_{2}^{B})\op{1}
\end{equation}
in the distributional sense smeared with any $\bar{F}_{1}$ and $\bar{F}_{2}$ on $\hor$.

We now briefly present the quantization of $\Alg$ presented in sec.~\ref{sec:HilbmmKG} from the point of view of the horizon quantization presented in this subsection. By de Sitter invariance, $\Omega_{\BD}$ has zero memory and, as we argued previously, \eqref{eq:qPiF} is an obstruction to $\Omega_{\BD}$ being a normalizable state on $\Alg$. However, $\Omega_{\BD}$ is a well-defined, Gaussian state on the shift-invariant algebra generated by $\op{\Pi}(\bar{F})$ with two-point function
\begin{equation}
\label{eq:OBDhor}
\braket{\Omega_{\BD}|\op{\Pi}(x_{1})\op{\Pi}(x_{2})|\Omega_{\BD}} = -\frac{1}{\pi}\frac{\delta_{\bb{S}^{2}}(x_{1}^{A},x_{2}^{B})}{(V_{1}-V_{2}-i0^{+})^{2}}
\end{equation}
smeared with any $\bar{F}_{1},\bar{F}_{2}$. The vacuum $\Omega_{\BD}$ yields a one-particle Hilbert space $\Hilb_{0}$ with inner product 
% \note{put angular integral sign here and other places!!} 
\begin{equation}
\braket{\bar{F}_{1}|\bar{F}_{2}}_{0} = 2\int_{0}^{\infty}\omega d\omega ~\int_{\bb{S}^{2}}d\Omega_{2}~\bar{F}_{1}(\omega,x^{A})\bar{F}^{\ast}_{2}(\omega,x^{A})
\end{equation}
where $\bar{F}(\omega,x^{A})$ is the Fourier transform of $\bar{F}(V,x^{A})$ with respect to $V$ and $\bar F^{\ast}$ is the complex conjugate. The corresponding Fock representation $\Fock_{0}$ is an irreducible representation of $\Alg_{\NCM}$ with vanishing memory. The construction of the memory Fock representation is obtained by applying the shift $\op{\Pi} \to \op{\Pi} + \partial_{V}\Xi_{q}\op{1}$  where $\Delta[\Xi_{q}] = q$. It is straightforward to show that this map is unitarily implementable on $\Fock_{0}$ if and only if the norm
\begin{equation}
\label{eq:Vqnorm}
\norm{\Xi_{q}}_{0}^{2} = 2\int_{0}^{\infty}\omega d\omega ~\int_{\bb{S}^{2}}d\Omega_{2}~\abs{\Xi_{q}(\omega,x^{A})}^{2}
\end{equation}
on $\Hilb_{0}$ is finite \cite{Ashtekar:1981sf,PSW-IR}. Since $\Delta[\Xi_{q}]=q$, it follows that the Fourier transform diverges at low frequencies as $\Xi_{q}(\omega,x^{A})\sim q/\omega$ and so \cref{eq:Vqnorm} diverges logarithmically at low frequencies for $q\neq 0$. Thus, the states with non-vanishing memory $q$ contain an infinite number of soft radiative quanta and lie in the memory Fock space $\Fock_{q}$ which is unitarily inequivalent to $\Fock_{0}$. The full representation of the algebra $\Hilb_{\dS}$ is obtained by direct integration over $\Fock_{q}$ as in \cref{sec:HilbmmKG}. The explicit action of the observable $\op{\Pi}(F)\in \Alg$ on $\Hilb_{\dS}$ where $q[F]\neq 0$ can be obtained in an analogous manner to the general action of $\op{p}$ on $\Hilb_{\dS}$ obtained in appendix \ref{sec:covariant-stuff}.

\begin{remark}[{\em Existence and non-existence of states invariant under de Sitter subgroups}]
States invariant under subgroups of de Sitter have been previously investigated by a number of authors \cite{Allen:1985ux,Kay_1988,Allen:1987tz,Kirsten:1993ug}. In this remark we rederive and, in some cases, extend these results in a simple and straightforward way utilizing the quantization presented in the previous subsections. The purpose of this remark is to illustrate that while there is, at most, one de Sitter invariant state, the requirement of invariance under subgroups of de Sitter is either far too restrictive and yields no proper states or is not restrictive enough and admits an uncountably many such states. Thus, in this regard, considering states invariant under any subgroup of de Sitter does not appear to have the same utility as considering de Sitter invariant states.

First, let us consider the subgroup \(O(4)\), which preserves the global foliation by Cauchy surfaces $\Sigma \cong \bb{S}^{3}$ (see \cref{sec:dS}). In this choice of foliation, the quantization of the massless scalar field is explained in \cref{sec:MMC}. Since the group $O(4)$ does not ``mix modes'' in \cref{eq:modeglobal}, the mode $\xi(t)$ with unit conserved charge is invariant under this group and the choice of vacua $\ket{\Omega_{q\xi}}$ are also invariant. With this choice of vacua, $O(4)$ maps each fibre $\Fock_{q}$ in the direct integral \cref{eq:Hilbds} into itself. It follows that any state in $\Hilb_{\dS}$ of the form 
\be\label{eq:PhiO4}
    \ket{\Phi} \defn \int_{\bb{R}}dq~\psi(q)\ket{\Omega_{q\xi}}
\ee
where $\psi(q)$ is a square-integrable function on $\bb{R}$, is invariant under $O(4)$. Thus, there are an uncountably infinite number of $O(4)$-invariant states. When the wavefunction \(\psi(q)\) is chosen to be a Gaussian then the states \cref{eq:PhiO4} are equivalent to the $2$-parameter family of $O(4)$-invariant Gaussian states obtained by Allen and Folacci \cite{Allen:1987tz}.

Another subgroup considered by Allen is \(E(3)\), the group of isometries of flat Euclidean \(\bb R^3\), which preserves the foliation of de Sitter by flat slices \cite{Allen:1985ux}. The vector fields generating this subgroup are spacelike in the Poincar\'e patch and are null on the cosmological horizon. In terms of the affine parameter \(V\) of the cosmological horizon, the transformations in \(E(3)\) correspond to \(V \to V + c_{m}Y_{1m}(x^{A})\), a shift by any $\ell=1$ spherical harmonic along with rotations of the \(2\)-sphere cross-sections of the horizon. The states in $\Hilb_\dS$ on the horizon which are invariant under this subgroup can be explicitly constructed on each Fock space $\Fock_{q}$ following \cite{Prabhu:2024lmg} --- where they were denoted by \(\ket{\varnothing; q}\) --- and, have vanishing spatial momentum with charge \(q\). However, it was shown in \cite{Prabhu:2024lmg} that all such states are improper (non-normalizable) states, consistent with the conclusions of \cite{Allen:1987tz}. Integrating these states over $q$ does not make the states normalizable.

Lastly, consider the transformation generated by the horizon Killing field given by \(V \to e^{\kappa c}V\) where \(c\) is a constant. The only state invariant under this transformation is the Bunch-Davies state $\Omega_{\BD}$ \cite{Kay_1988}, which is an improper state. Thus, there are no proper states invariant under any subgroup of the de Sitter group containing the horizon Killing isometry.
\end{remark}

%%---------------------------------------------------------------------------------
\subsection{Quantization of the massive scalar}
\label{subsec:mscalar}
The algebra $\Alg_{m}$ of the massive scalar field is generated by the smeared observables $\op{\phi}(x)$ which satisfy the commutation relations \eqref{eq:commKG} where $E$ is now the advanced-minus-retarded Greens function for the massive theory and 
\begin{equation}
\label{eq:mphieq}
(\Box - m^{2})\op{\phi}(x) = 0 
\end{equation}
in the distributional sense. The quantization algebra on the cosmological horizon is identical to the algebra of horizon observables $\op{\Pi}(F)$ where $F=Ef\vert_{\hor}$. Indeed, the horizon observable is equivalent to the local field observable (i.e., $\op{\Pi}(F) = -\tfrac{1}{2}\op{\phi}(f)$ for any $f$), satisfies the commutation relations \eqref{eq:Picomm} and the symplectic product is again given by \eqref{eq:sympprodhor}. Finally, the memory $\op{\Delta}$ is given by \eqref{eq:memscalar} and controls the infrared behavior of quantum fields. 

The key difference between the massless and massive scalar is that, in the latter case, solutions $Ef$ exponentially decay at early and late times \cite{Dappiaggi:2008dk}. Thus the solutions asymptotically decay on the horizon and, by \eqref{eq:memphidiff}, the horizon observables satisfy 
\begin{equation}
[\op{\Delta}(\lambda),\op{\Pi}(F)] = - \frac{1}{2} [\op{\Delta}(\lambda),\op{\phi}(f)] = 0
\end{equation}
for any $\lambda(x^{A})$ and any $F=Ef$. Consequently, there is no obstruction to the existence\footnote{There exist a one-parameter family of de Sitter invariant states 
$\Omega_{\alpha}$ known as ``$\alpha$-vacua'' where $\Omega_{\BD}$ is the $\alpha=0$ state \cite{Allen:1985ux,Mottola:1984ar,Bousso:2001mw,deBoer:2004nd}. However, for $\alpha\neq 0$, the states $\Omega_{\alpha}$ are not Hadamard and therefore cannot be defined on the algebra of nonlinear observables. Indeed, it has been shown that the (locally smeared) renormalized stress tensor has infinite fluctuations in all $\alpha$-vacua states with $\alpha\neq 0$ \cite{Brunetti:2005pr}.} of $\Omega_{\BD}$ which is a normalizable, Gaussian state on $\Alg_{\KG,m}$ with vanishing one-point function and two-point function given by \eqref{eq:OBDhor}. The irreducible representation of the full algebra containing $\Omega_{\BD}$ is the zero memory Fock space $\Fock_{0}$.

This result is in complete agreement with the standard construction of $\Fock_{0}$ on a compact spatial slice, e.g. \cite{Allen:1985ux}. Indeed, the Hadamard nature of $\Omega_{\BD}$ in the UV ensures the existence of a dense set of Hadamard states in $\Fock_{0}$ and the above argument indicates that there are no ``IR modes'' which are not normalizable in $\Fock_{0}$. Nevertheless, it is 
reasonable to ask whether one should consider states with non-vanishing horizon memory for the free, massive scalar field. Such states lie in Fock representations $\Fock_{\Delta}$ which are unitarily inequivalent to $\Fock_{0}$. On a compact spatial slice, the lack of any IR modes indicates that the UV behavior of states with memory does not approach the UV behavior of $\Omega_{\BD}$. Therefore, states with memory cannot be Hadamard and are therefore unphysical in the free theory. The analysis in the following sections suggests that similar conclusions can be drawn for states of the free electromagnetic field with electromagnetic memory as well as states of the free linearized gravitational field with gravitational memory. As we will see in sec.~\ref{sec:sources}, physical states with memory can be produced in the presence of sources.

\subsection{Quantization of the free electromagnetic field}
\label{subsec:EM}

We now consider the quantization of the electromagnetic field in de Sitter spacetime. The basic degree of freedom that we must quantize is the vector potential $A_{a}(y)$ which satisfies the Maxwell's equation
\begin{equation}
\label{eq:MWell}
\Box A_{a} - \nabla^{b}\nabla_{a}A_{b} = 0.
\end{equation}
Any two solutions of \eqref{eq:MWell} are physically equivalent if they can be related by the gauge transformation 
\begin{equation}
A_{a} \mapsto A_{a} + \nabla_{a}\lambda.
\end{equation}

In the quantum theory, the algebra of observables $\Alg_{\EM}$ is generated by the smeared vector potential 
\begin{equation}
\op{A}(f) \defn \int_{M}\sqrt{-g}d^{4}y~\op{A}_{a}(y)f^{a}(y)
\end{equation}
where $f^{a}$ is a test tensor on $M$ and $\op{A}_{a}$ satisfies Maxwell's equations \eqref{eq:MWell} in the distributional sense. However, $\op{A}_{a}$ is gauge dependent and the necessary and sufficient condition to eliminate this gauge dependence is to restrict the class of smearing functions to divergence-free test vector field (i.e., $f^{a}$ satisfy $\nabla_{a}f^{a}=0$) \cite{Wald_1995}. Under this restriction, the observable $\op{A}(f)$ is gauge invariant. When smeared with divergence-free test vector fields, the retarded and advanced Green's functions \(E_{ab}^\pm(y,y')\) are well-defined \cite{Pfenning:2009nx}. The algebra generated by $\op{A}(f)$ is subject to the commutation relations 
\begin{equation}
[\op{A}_{a}(y),\op{A}_{b}(y^{\prime})] = i E_{ab}(y,y^{\prime})\op{1}
\end{equation}
where, as before, $E_{ab}$ is the advanced-minus-retarded Green's function and is well-defined in the distributional sense smeared with any (divergence-free) test tensors $f_{1}^{a}$ and $f_{2}^{b}$. This covariant quantization can be directly related to the canonical quantization of the initial data on any Cauchy surface exactly as in the scalar field case, discussed above. 

Note that, in contrast to the massless scalar, the electromagnetic field in de Sitter spacetime does not have a conserved charge which controls the late time behavior of (source-free) solutions.
However, as in the case of the scalar field, we can define an \emph{electromagnetic horizon memory} observable on any cosmological horizon. While the algebra \(\Alg\) in the spacetime is fully gauge invariant, to describe the quantization on a horizon $\hor$ it is convenient to choose a gauge where the component of the vector potential along the affine generators is zero, i.e., \(A_V = 0\) on $\hor$. Let the pullback of $A_{a}$ to $\hor$ be $A_{A}(V,x^{A})$, where capital Latin indices denote angular components on the horizon. The symplectic form on the horizon is given by 
\be
    \Omega_{\hor}(A_{1},A_{2}) = \int_{\hor} dVd\Omega_{2}~(A_{1A}E_{2}^{A} - A_{2A}E_{1}^{A})
\ee
where $E_{A} = -\partial_{V}A_{A}$ is the ``electric field'', i.e., the \(F_{AV}\) component of the Maxwell field tensor at \(\hor\), which characterizes the electromagnetic radiation through the horizon. Given any solution $F_{A}$ on the horizon, we define the observable 
\be
\label{eq:Esymp}
\op{E}(F) \defn -\frac{1}{2} \Omega_{\hor}(\op{A},F) = -\frac{1}{2} \op{A}(f)
\ee
where $F_{a}\defn(Ef)_{a}$ on $\hor$. We label the observable as ``$\op{E}$'' because for any $F$ that decays at asymptotically early and late affine times, \eqref{eq:Esymp} is $-\partial_{V}\op{A}_{A}$ smeared with $F$ on $\hor$. If $F$ has memory, the solution cannot simultaneously decay at early and late times, however we will still label the horizon observable as $\op{E}$.  

The analog of the scalar horizon memory in the electromagnetic case is given by
\be
\label{eq:mem}
    \Delta_A(x^B) &\defn \lim_{V\to +\infty}A_A(V , x^B) - \lim_{V\to -\infty}A_A(V, x^B) \\
    &= - \int_{-\infty}^\infty dV~ E_A(V,x^B)
\ee
which is manifestly gauge invariant. The smeared memory observable is given by
\be
    \Delta(\lambda)  = \int_{\mathbb{S}^{2}} d\Omega_2~ \Delta_A(x^B) \ms D^A \lambda(x^B) = \Omega_{\hor}(A, \delta_{\lambda}A) 
\ee
where $\delta_{\lambda}A_{A}\defn \ms{D}_{A}\lambda(x^{A})$ is a pure gauge solution on $\hor$. Since $\op{\Delta}(\lambda)$ generates gauge transformation and $\op{A}(f)$ is gauge-invariant we have that
\begin{equation}
[\op{\Delta}(\lambda),\op{A}(f)]=0.
\end{equation}
Thus, in contrast to the massless scalar, all of the observables commute with the memory and so there is no obstruction to the existence of the Bunch-Davies vacuum. 

Additionally, this commutator\footnote{As explained below eq.~\ref{eq:Esymp}, $\op{E}(F)$ is only the gauge invariant electric field if $F$ sufficiently decays at asymptotically early and late affine times.} can be expressed in terms of the memory of the source-free solution $Ef$ generated by $f$
\begin{equation}
[\op{\Delta}(\lambda),\op{A}(f)] =  -2 \Omega_{\mc{H}}(\delta_{\lambda}A,F)\op{1} = 2i\Delta^{(F)}(\lambda)\op{1}
\end{equation}
where $\Delta^{(F)}$ is the memory of the solution $F_{A}$ smeared with $\lambda(x^{A})$. Thus, the fact that memory commutes with all local gauge invariant observables directly implies that all solutions $Ef$ must decay at asymptotically early and late times. 

This result equivalently follows from the source-free Maxwell equations that the solutions $(Ef)_{a}$ exponentially decay in the past and future. To see this we note that Maxwell's equations restricted to the horizon yield 
\be
\label{eq:MXwelleqhor}
    \ms D^A E_A = \partial_{V} E_r 
\ee
where \(E_r \defn F_{ab}\ell^{a}n^{b}\vert_{\hor} \) is the electric field component along a null ``radial'' direction $\ell^{a}=(\partial/\partial r)^{a}$ transverse to the horizon (i.e. if $n^{a}=(\partial/\partial V)^{a}$ is a null generator of the horizon then $\ell^{a}$ satisfies $\ell^{a}n_{a}=1$). Integrating \eqref{eq:MXwelleqhor} along the null generators of horizon and using \eqref{eq:mem} we obtain\footnote{This component of Maxwell's equation on the horizon yields a formula for the ``electric parity'' memory. The ``magnetic parity'' memory satisfies $\epsilon^{AB}\ms{D}_{A}\Delta_{B}= B_{r}\big\vert^{V=+\infty}_{V=-\infty} $ where $B_{r}$ is the radial component of the magnetic field on the horizon \cite{Gralla:2023oya}. The magnetic parity memory does not generate gauge transformations however. by similar arguments as presented in sec.~\ref{subsec:EM}, $B_{r}$ exponentially decays for all smooth solutions with initial data of compact support. } 
\be\label{eq:EMconst1}
    \ms{D}^{A}\Delta_{A} = \lim_{V\to \infty}E_{r}(V,x^{A}) - \lim_{V\to - \infty}E_{r}(V,x^{A}). 
\ee
Thus, in the source-free case, the horizon memory depends on the early and late time behavior of the radial electric field. We now show that, {for smooth solutions}, $E_{r}$ decays exponentially at early and late times. This follows from the fact that the free Maxwell's equations are conformally-invariant, and thus, the field tensor \(F_{ab}\) of any smooth solution in de Sitter will extend smoothly to the conformal boundaries \(\scri^\pm\). It also follows from the conformal form of the de Sitter metric (\cref{eq:dS-metric-conf}) that the null vectors \((\csc\eta)n^{a}  \) and \((\csc\eta ) \ell^{a} \) have smooth limits to \(\scri^\pm\) along the horizon. Thus, $E_{r} \sim \sin^{2}\eta$ in conformal time and so we have that 
\begin{equation}
\label{eq:Er}
E_{r} \sim e^{-|t|/L}
\end{equation}
as the global time $t\to \pm \infty$. As
the radial electric field exponentially decays at early and late times. Therefore, $\Delta^{(F)}$ vanishes for any $F=Ef$. The two arguments, gauge invariance and decay of fields, are connected. Simply by gauge invariance\footnote{The argument that gauge invariance implies decay of the fields is independent of conformal invariance and holds in any dimension.} and \eqref{eq:EMconst1}, we can conclude that the fields must decay.

Consequently, there is no obstruction to the existence $\Omega_{\BD}$ in the electromagnetic case and, indeed, one can straightforwardly construct it. The de Sitter invariant vacuum is the Gaussian state with vanishing one-point function and two-point function given by
\begin{equation}
\braket{\Omega_{\BD}|\op{E}_A(x_{1})\op{E}_B(x_{2})|\Omega_{\BD}} = - \frac{1}{4\pi} \frac{q_{AB}\delta_{\bb{S}^{2}}(x_{1}^{A},x_{2}^{A})}{(V_{1}-V_{2}-i0^{+})^{2}}
\end{equation}
smeared with any $F_{1},F_{2}$ \cite{Allen:1985wd}. By eq.~\ref{eq:Esymp}, this initial data uniquely determines the global quantum state. Furthermore the irreducible representation of $\Alg_{\EM}$ containing $\Omega_{\BD}$ is the zero memory Fock space $\Fock^{\EM}_{0}$.

%%==============================================================
\subsection{Quantization of the free linearized gravitational field}
\label{subsec:GR}

We conclude this section by considering the quantization of the linearized gravitational field on a de Sitter background. Much of the analysis parallels the discussion of the massive scalar and the electromagnetic field, so we will primarily focus on the key differences which arise in quantizing the graviton field. The basic degree of freedom is the metric perturbation $\gamma_{ab}(y)$ which satisfies 
\begin{equation}
\label{eq:lingrav}
\Box_{g}\gamma_{ab} +\nabla_{a}\nabla_{b}\gamma^{c}{}_{c}-\nabla_{c}\nabla_{a}\gamma^{c}{}_{b}-\nabla_{c}\nabla_{b}\gamma^{c}{}_{a}+2\Lambda \gamma_{ab}=0
\end{equation}
and diffeomorphisms of the full metric result in a gauge freedom of the perturbation given by 
\begin{equation}
\gamma_{ab} \mapsto \gamma_{ab} + 2\nabla_{(a}\xi_{b)}
\end{equation}
In the quantum theory, the algebra of observables $\Alg_{\GR}$ is generated by the smeared metric perturbation 
\begin{equation}
\op{\gamma}(f) \defn \int_{M}\sqrt{-g}d^{4}y~\op{\gamma}_{ab}(y)f^{ab}(y)
\end{equation}
where $f^{ab}$ is a symmetric test tensor on $M$ and $\op{\gamma}_{ab}$ satisfies the linearized Einstein equations in the distributional sense. The necessary and sufficient condition to eliminate the gauge dependence of $\op{\gamma}$ is to restrict to divergence-free test tensors (i.e., $f^{ab}$ satisfying $\nabla_{a}f^{ab}=0$). Under this restriction, $\op{\gamma}(f)$ is gauge-invariant and the retarded and advanced Green's functions $E_{abcd}^{\pm}$ are well defined \cite{Wald_1995,Fewster:2012bj}. The covariant commutation relations are given by 
\begin{equation}
[\op{\gamma}_{ab}(y),\op{\gamma}_{cd}(y^{\prime})] = i E_{abcd}(y,y^{\prime})\op{1}
\end{equation}
where $E_{abcd}$ is the advanced-minus-retarded propagator. 

As in the case of the massive scalar and electromagnetic field, the gravitational field does not admit a conserved charge that controls the infrared behavior of solutions. However, the theory does admit a gauge-invariant {\em gravitational horizon memory} observable on any cosmological horizon. To describe the quantization on the cosmological horizon it is useful to choose the gauge $n^{a}\gamma_{ab}\vert_{\mc{H}} = 0 =q^{ab}\gamma_{ab}\vert_{\mc{H}}$ where $n^{a}$ is the horizon generator and $q^{ab}$ is an inverse metric on the horizon. This gauge condition ensures that horizon location is not changed due to the first order perturbation \cite{Hollands:2012sf} and, in  this gauge, the initial data on the horizon is encoded in the angular components $\gamma_{AB}$ with symplectic form
\begin{equation}
\Omega_{\mc{H}}(\gamma_{1},\gamma_{2}) = \int_{\mc{H}}~dVd\Omega_{2}~(\gamma_{1AB}\sigma_{2}^{AB} - \gamma_{2AB}\sigma_{1}^{AB})
\end{equation}
where $\sigma_{AB}\defn \frac{1}{2}\partial_{V}\gamma_{AB}$ is the perturbed shear of the horizon. Given any solution $F_{AB}=(Ef)_{ab}\vert_{\mc{H}}$ with source $f_{ab}$ we define the observable 
\begin{equation}
\label{eq:sigma}
\op{\sigma}(F)\defn -\frac{1}{2} \Omega_{\mc{H}}(\op{\gamma},F) = -\frac{1}{2} \op{\gamma}(f)
\end{equation}
which, if $F$ decays at asymptotically early and late times, is $\frac{1}{2}\partial_{V}\op{\gamma}_{AB}$. Finally, the analog of the memory in the gravitational case is given by 
\begin{align}
\Delta_{AB}\defn& \frac{1}{2}\bigg[\lim_{V\to \infty}\gamma_{AB}(V,x^{A}) - \lim_{V\to -\infty}\gamma_{AB}(V,x^{A})\bigg] \nonumber \\
=&\  \int_{\bb{R}}dV~\sigma_{AB}(V,x^{A})
\end{align}
which we note is gauge invariant. We define the smeared memory observable as 
\begin{equation}
\Delta^{\GR}(\lambda) \defn  \int_{\bb{S}^{2}}d\Omega_{2}~\Delta_{AB}(x^{C})\ms{D}^{A}\ms{D}^{B}\lambda(x^{C}) = -\Omega_{\mc{H}}(\gamma,\delta_{\lambda}\gamma)
\end{equation}
where, the initial data in the second slot is
\begin{equation}
\delta_{\lambda}\gamma_{AB} = \bigg(\ms{D}_{A}\ms{D}_{B}-\frac{1}{2}q_{AB}\ms{D}^{2}\bigg)\lambda(x^{C})
\end{equation}
is pure gauge initial data and corresponds to an infinitesimal supertranslation $V\to V+\lambda(x^{C})$. The commutator of the memory with the local observable is again determined by the ``memory'' of the observable 
\begin{equation}
\label{eq:memGRcomm}
[\op{\Delta}^{\GR}(\lambda),\op{\gamma}(f)] = -i\Delta^{(F),\GR}(\lambda)\op{1}
\end{equation}
where $\Delta^{(F),\GR}$ is the memory of solution $(Ef)_{ab}$ on $\hor$. $\op{\Delta}(\lambda)$ generates supertranslations and since $\op{\gamma}(f)$ is a local, gauge invariant observable it must commute with the memory observable. Thus, as in the electromagnetic case, we immediately see that there is no obstruction to the existence of a Bunch-Davies vacuum.  

We can also see this by analyzing the decay of the linearized solutions $(Ef)_{ab}$. Consider the linearized Bianchi identity\footnote{\label{foot:Bianchi}This is obtained from combining the two equations $\ms{D}^{A}E_{AB}=-\partial_{V}E_{Br}$ and $\ms{D}^{A}E_{Ar} =- \partial_{V}E_{rr}$ on the horizon \cite{Danielson:2022tdw,Danielson:2022sga} and using the fact that $E_{AB}=-\tfrac{1}{2}\partial_{V}\sigma_{AB}$.} for the linearized, electric Weyl tensor $E_{ac} = C_{abcd}n^{b}n^{d}\vert_{\hor}$ on the horizon \cite{Danielson:2022tdw,Danielson:2022sga}
\begin{equation}
\label{eq:Bianchisourcefree}
\ms{D}^{A}\ms{D}^{B}E_{AB} = \partial_{V}^{2}E_{rr}
\end{equation}
where $E_{AB}=-\partial_{V}\sigma_{AB}$ and $E_{rr} \defn C_{abcd}\ell^{a}n^{b}\ell^{c}n^{d}\vert_{\hor}$. Integrating this equation twice with respect to affine time yields 
\begin{equation}
-\ms{D}^{A}\ms{D}^{B}\Delta_{AB} = \lim_{V\to \infty}E_{rr}(V,x^{A}) - \lim_{V\to -\infty}E_{rr}(V,x^{A})
\end{equation}
Therefore, in the source-free case, the horizon memory depends upon the early and late time or the radial component of the electric Weyl tensor. However, the radial electric Weyl tensor decays exponentially; this follows from the analysis of Friedrich in nonlinear general relativity \cite{Friedrich:1986qfi}. Consequently, the memory vanishes for all source-free solutions and there is no obstruction to the existence of a de Sitter invariant vacuum. We emphasize that the lack of memory is a gauge invariant property of the solutions. The de Sitter invariant state is the Gaussian state with vanishing one-point function and two-point function on $\hor$ given by  
\begin{equation}
\braket{\Omega_{\BD}|\op{\sigma}_{AB}(x_{1})\op{\sigma}_{CD}(x_{2})|\Omega_{\BD}} = \frac{1}{8\pi} \frac{q_{AC}q_{BD}\delta_{\bb{S}}(x_{1}^{A},x_{2}^{A})}{(V_{1}-V_{2}-i0^{+})^{2}}
\end{equation} 
smeared with any (trace-free) solutions $F_{1,AB}$ and $F_{2,CD}$. By eq.~\ref{eq:sigma}, this initial data uniquely determines the global quantum state. Furthermore, the irreducible representation containing $\Omega_{\BD}$ is the zero memory Fock space $\Fock_{0}^{\GR}$. 

\section{Infrared Particle Production in the Presence of Sources}
\label{sec:sources}
In the previous section, we showed that memory commutes with the algebra of local observables of the free massive scalar, free electromagnetic field and free, linearized gravitational field. Therefore, one can construct a (zero memory) Fock space of states which decay at asymptotically early and late times with a normalizable Bunch-Davies vacuum. By contrast, for the massless scalar field, the constant memory does not commute with the local observables. The quantum states on this algebra contain an infinite number of soft radiative quanta and do not decay at asymptotically early and late times. 

In this section, we analyze the infrared behavior of quantum states in the presence of a ``source.'' We will focus primarily on the electromagnetic case where the source is a charge-current $j^{a}$ and simply state the analogous analysis and conclusions for the scalar and gravitational cases at the end of this section. Since de Sitter spacetime is spatially closed, Maxwell's equations implies that $j^{a}$ must have zero total charge 
\begin{equation}
\int_{\Sigma}\sqrt{h}d^{3}x~n^{a}j_{a}=0
\end{equation}
on any Cauchy surface $\Sigma$. Furthermore, Maxwell's equations on any cosmological horizon now aquire an additional term\footnote{There is also an evolution equation that governs the behavior of magnetic fields on the horizon (see, e.g., \cite{Gralla:2023oya}). We will focus, for simplicity, on the behavior of electric fields but the same analysis will be applicable to magnetic fields and the production of ``magnetic parity memory'' \cite{Satishchandran:2019pyc}} 
\begin{equation}
\label{eq:EA}
\ms{D}^{A}E_{A} = \partial_{V}E_{r} + j_{V}
\end{equation}
where, again, $E_{A}=-\partial_{V}A_{A}$ and $j_{V}$ is the charge-current flux through the cosmological horizon. Integrating eq.~\ref{eq:EA} with respect to affine time yields 
\be
\label{eq:memeqhorj}
\ms D^A \Delta_A = E_r\bigg\vert_{V=-\infty}^{V=+\infty} + \int_{-\infty}^{\infty}dV~j_{V}
\ee
Both terms are generally non-vanishing. The second term is non-vanishing whenever the charge-current crosses the horizon. As we will see, $j^{a}$ creates a non-vanishing radial Coulomb field on the horizon so the first term is generally non-vanishing if the source persists to future or past infinity. Thus, quite generally, the vector potential $A_{A}$ will fail to decay at asymptotically early and late times. 

We now analyze the relationship between these infrared tails and production of soft radiation in the quantum theory. To fully analyze this problem one should consider the full interacting theory of QED in de Sitter spacetime with a quantized charged source. We will instead analyze the much simpler and tractable problem of a quantum electromagnetic field coupled to a classical source. We thereby neglect the fluctuations of the charge current operator which can be approximated as $\op{j}^{a}=j^{a}\op{1}$. This theory has the advantage of being completely soluble and is essentially the same approximation originally used by Bloch and Nordseik to investigate the infrared behavior\footnote{More precisely the authors of \cite{Bloch:1937pw} obtained this approximation from considering the ``low-frequency'' or ``large mass'' limit (i.e. $\omega/m\ll 1$) of QED where $m$ is the mass of the electron and $\omega$ is the frequency of the photon.} of quantum fields in flat spacetime \cite{Bloch:1937pw}.

The computation of the quantized radiation emitted by a classical current $j^{a}$ in any spacetime with a bifurcate Killing horizon was previously analyzed by one of us together with D. Danielson and R. M. Wald \cite{Danielson:2022sga}. We now give a brief review of their analysis and results focusing on the key points which are most salient to this paper. Since we are interested in the production of infrared radiation emitted by $j^{a}$ we will our restrict attention to charge currents which vanish in the asymptotic past and future. More precisely, we will first consider charge-currents such that there exists a global times $t_{0}<t_{1}$ such that $j^{a}(t,x^{i})$ vanishes for all $t<t_{0}$ and $t>t_{1}$. We will first consider the radiation emitted by such currents of compact support in spacetime. We will then consider the limit as $t_{1}\to \infty$.  

For $t<t_{0}$, $j^{a}$ vanishes, and so the vector potential $\op{A}_{a}$ satisfies the source-free Maxwell equations at early times. In the Heisenberg representation we will denote the vector potential at early times as
\begin{equation}
\op{A}^{\inn}_{a}(t,x^{i})= \op{A}_{a}(t,x^{i}) \textrm{ for any $t<t_{0}$.}
 \end{equation}
The initial state of the electromagnetic field may be specified by giving the ``radiation state'' of $\op{A}^{\inn}$. We assume that the initial quantum state is in the Bunch-Davies vacuum. For $t\geq t_{0}$, the source is non-vanishing and the solution for $\op{A}_{a}$ in the Heisenberg picture with source $\op{j}=j\op{1}$ is \cite{Yang:1950vi}
 \begin{equation}
\op{A}_{a} = \op{A}^{\inn}_{a} +(E^{+}j)_{a}\op{1}
 \end{equation}
 where $(E^{+}j)_{a}$ is the classical retarded solution to Maxwell's equation with source $j^{b}$. The correlation functions of the quantum state $\Psi$ resulting from this interaction are given by %\note{the retarded Green's function notation below should also be fixed}
 \be
&\quad \braket{\Psi|\op{A}_{a_{1}}(y_{1})\dots \op{A}_{a_{n}}(y_{n}) |\Psi} \\
&=\bra{\Omega_{\BD}} [\op{A}^{\inn}_{a_{1}}(y_{1})+(E^{+}j)_{a_{1}}(y_{1})\op{1}] \\
&\qquad\qquad \dots [\op{A}^{\inn}_{a_{n}}(y_{n})+(E^{+}j)_{a_{n}}(y_{n})\op{1}] \ket{\Omega_{\BD}}
 \ee
Similarly, for $t>t_{1}$, the retarded solution $E^{+}_{ab}[j^{b}]$ is a source free solution and so the quantum state $\ket{\Psi}$ at any time to the future of the support of $j^{a}$ is a free field state given by 
\begin{equation}
\ket{\Psi} = \exp(i\op{A}^{\textrm{in}}(j))\ket{\Omega_{\BD}}
\end{equation}
It was shown in  \cite{Danielson:2022sga} (see also \cite{Gralla:2023oya,Danielson:2024yru}) that the total expected number of photons emitted by the source $j_{a}$ is given by one-particle norm of the advanced-minus-retarded solution $(Ej)_{a} = (E^{-}j)-(E^{+}j)_{a}$ which is the unique, global free field solution which satisfies $(Ej)_{a}=(E^{+}j)_{a}$ for $t>t_{1}$. Thus we have that  
\begin{equation}
\label{eq:N}
\braket{N}_{\Psi} = \braket{\Omega_{\BD}|\op{A}^{\rm in}(j)^{2}|\Omega_{\BD}}= \norm{ Ej}_{\Sigma}^{2}
\end{equation}
where we note that, since the norm is conserved on source-free solutions, $\Sigma$ is {\em any} Cauchy surface in the spacetime. Furthermore, the necessary and sufficient condition for the radiation state $\ket{\Psi}$ to lie in the standard Fock space $\Fock_{0}$ is that $\braket{N}_{\Psi}<\infty$.

The radiation emitted by $j^{a}$ is controlled by the behavior of the source free solution $(Ej)_{a}$. Taking $\Sigma$ be any cosmological horizon $\mc{H}$, the norm given by eq.~\ref{eq:N} can be expressed in terms of the vector potential as
\begin{equation}
A_{a}(V,x^{A})\defn (Ej)_{a}\bigg\vert_{\mc{H}}
\end{equation}
which is the pullback of the solution to the horizon. On any cosmological horizon the norm is 
\begin{equation}
\label{eq:cosmnorm}
\norm{ Ej] }_{\mc{H}}^{2} =\frac{1}{2}\int_{\bb{S}^{2}}d\Omega \int_{0}^{\infty}~\omega  d\omega~q^{BC}\overline{\hat{A}_{B}(\omega,x^{A})}\hat{A}_{C}(\omega,x^{A})
\end{equation}
where $\hat{A}_{A}$ is the Fourier transform of $A_{A}$ with respect to affine time. 
In section~\ref{subsec:EM} we showed that for any smooth $j^{a}$ of compact support, the solution $(Ej)_{a}$ decays at asymptotically early and late times. In particular, this implies that memory of $(Ej)_{a}$ vanishes on any cosmological horizon. Thus, the expected number of photons is finite and $\Psi\in\Fock_{0}$. 

However, it was shown in \cite{Danielson:2022tdw,Danielson:2022sga} that if the source persists for sufficiently long times relative to the Hubble time then the norm of $(Ej)_{a}$ can be large and thereby emits a large number of soft photons.  In particular, in the limit as the source persists to infinity, it emits an infinite number of soft photons. To illustrate this, consider an inertial point dipole which is smoothly created at $t=t_{0}$ and smoothly destroyed at $t=t_{1}$. An inertial source in de Sitter spacetime defines a ``static patch'' with future and past cosmological horizons. In terms of static coordinates $(\tau,\vec{x})$ in the patch, the current is explicitly given by 
\begin{equation}
j_{\rm{dip.}}^{a} =\frac{2}{\sqrt{g}}\tau^{[a}s^{b]}\nabla_{b}[f(\tau)\delta^{(3)}(\vec{x})]
\end{equation}
where $\tau^{a}=(\partial/\partial \tau)^{a}$ corresponds to the timelike Killing field of the patch, $s^{a}$ is the spatial unit vector corresponding to the direction of the dipole and $\delta^{(3)}(\vec{x})$ is a ``coordinate delta function'' defined so that $\int \delta^{(3)}(\vec{x}) d^{3}x = 1$. If $\tau_{1}$ and $\tau_{2}$ correspond to the global times $t_{1}$ and $t_{2}$ then the smooth function $f(\tau)$ satisfies 
\begin{equation}
f(\tau) =
\begin{cases}
p \eqsp \textrm{ for } |\tau| < \tfrac{T}{2} \\
%0 \eqsp \textrm{ for } |\tau| > \tfrac{T + \tilde{T}}{2} \\
0 \eqsp \textrm{ for } \tau < -\tfrac{T}{2} - \tau_{1} \textrm{ and } \tau> \tfrac{T}{2}+\tau_{2}
\end{cases}
\end{equation}
where $p$ is constant corresponding to the strength of the dipole, $T$ corresponds to the proper time that the dipole persists until it is destroyed. The particular choice of a dipole as opposed to a current with any other higher multipole structure will not be relevant to the conclusions of this section and is only meant to be considered as a specific, concrete example with zero total charge. We now sketch the estimate of the expected number of photons emitted by this source following \cite{Danielson:2022sga}. 

While we could evaluate eq.~\ref{eq:N} on any Cauchy surface in de Sitter spacetime we will find that the number of photons can be straightforwardly evaluated on the future horizon of the Poincaré patch using the approach presented in this paper. The calculation of $\braket{N}_{\Psi}$ on the past horizon was given\footnote{In \cite{Gralla:2023oya} the authors considered the number of photons emitted by a dipole source created in the exterior of a Kerr black hole. However, the number of photons was evaluated on the past horizon of the bifurcate Killing horizon. This computation is identical to the calculation of eq.~\ref{eq:N} on the past horizon.} in \cite{Gralla:2023oya}. The computation of $\braket{N}_{\Psi}$ by integrating the local vacuum fluctuations in the first equality of eq.~\ref{eq:N} was given in \cite{Danielson:2024yru}. On the future horizon of the static patch, we have that $(Ej)_{a}\big\vert_{\mc{H}}=(E^{+}j)_{a}\big\vert_{\mc{H}}$. The inertial source generates a long-range radial Coulomb field on the horizon 
\begin{equation}
E_{r}(V,x^{A}) =
\begin{cases}
E^{\rm{dip.}}_{r}(x^{A}) \textrm{ for }|V|<V_{T}/2 \\ \nonumber 
0 \textrm{ for }V <-\tfrac{V_{T}}{2} - V_{1} \textrm{ and }V >\tfrac{V_{T}}{2} + V_{2}
\end{cases}
\end{equation}
where $V_{1}$ and $V_{2}$ correspond to the affine times that the source is created and destroyed respectively. The affine time $V_{T}=e^{T/L}$ is the duration of affine time corresponding to the proper time $T$ and $E^{\rm{dip.}}_{r}(x^{A})$ is the stationary Coulombic field of the dipole whose magnitude is of order $|E^{\rm{dip.}}_{r}|\sim p/L^{3}$. Integrating eq.~\ref{eq:EA}, using the fact that $E_{A}=-\partial_{V}A_{A}$ and recalling that $(Ej)_{a}$ is a source-free solution so $j_{V}=0$, we see that the initial change in $E_{r}$ on the horizon implies a change in the vector potential $A_{A}(V,x^{A})$. Thus, for $V<-V_{T}/2+V_{2}$ the vector potential on the horizon behaves as though it ``has memory'' with magnitude 
\begin{equation}
|\Delta E_{r}| \sim p/L^{3}.
\end{equation}
This change in the vector potential persists for an affine time $V_{T}=e^{T/L}$ before returning back to its original value. As previously mentioned, if the vector potential truly had memory and so the change in the vector potential was permanent then the norm would be logarithmically divergent at low frequencies. If the change in the vector potential persists for a long but finite amount of time (i.e., $T\gg L$) then it was shown in \cite{Danielson:2022tdw,Danielson:2022sga} that the number of soft photons grows logarithmically in $V_{T}$ and thus linearly in $T$ 
\begin{equation}
\braket{N}_{\Psi} \sim \bigg(\frac{|\Delta E_{r}|^{2}}{L}\bigg)T
\end{equation}
For the precise formula of the number of radiative quanta with numerical factors we refer the reader to \cite{Gralla:2023oya}. Thus, if the source exists for times much longer than the Hubble time then it will emit a large number of soft radiative quanta. As $T\to \infty$, $E_{r}$ will not vanish at asymptotically late time and the solution $(Ej)_{a}$ will have memory $\Delta$ determined by eq.~\ref{eq:memeqhorj}. In this limit, the source emits an infinite number of radiative quanta and the radiation state $\ket{\Psi} \in \Fock_{\Delta}$ will lie in an inequivalent Hilbert space of photons. 

To summarize, to obtain the expected number of photons produced by a source $j^{a}$ one must compute the norm of the source-free, advanced-minus-retarded solution $(Ej)_{a}$. To illustrate the kinds of sources that can produce a large number of soft photons we considered, for definiteness, an inertial source and showed that it emits a growing number of soft photons which diverges in the limit as the source goes to infinity. If the current is replaced by a congruence of geodesic charges (i.e., $j^{a}$ corresponds to charge-current ``dust'') then such sources will similarly radiate soft photons on long timescales relative to the Hubble time. While a detailed, general analysis of the radiative content of $(Ej)_{a}$ for general sources $j^{a}$  is outside of the scope of this paper, this analysis suggests that, on timescales much longer than the Hubble time, such sources will generally emit a large number of soft radiative quanta. 

An exactly similar analysis holds for a ``source'' coupled to a massive scalar field
where as opposed to an inertial dipole one can simply consider an inertial scalar charge. In the perturbative gravitational case, the gravitational field ``sources itself'' at second order in perturbation theory and this source could similarly radiate soft gravitons. To avoid the complications of computing higher-order perturbative effects, we will instead analyze the impact of introducing a stress-energy source $T_{ab}$ for the linearized field (i.e. the right hand side of eq.~\ref{eq:lingrav} is replaced by $-16 \pi G_{\rm{N}} T_{ab}$). Integrating the linearized Bianchi identity on the horizon with a non-vanishing stress tensor now yields\footnote{This formula follows from the same manipulations of the Bianchi identity that led to \eqref{eq:Bianchisourcefree} outlined in footnote~\ref{foot:Bianchi} with now the linearized Ricci tensor determined by the stress tensor via Einstein's equation.} 
\begin{equation}
\label{eq:bianchi}
-\ms{D}^{A}\ms{D}^{B}\Delta_{AB} = \Delta E_{rr} + 8 \pi G_{\rm{N}}  \int dV~ T_{VV}
\end{equation}
which is similar to eq.~\ref{eq:Bianchisourcefree} where now the right hand side receives an additional contribution due to the stress energy flux $T_{VV}$ through the horizon. As in the electromagnetic case, any massive body can produce a non-vanishing memory on any horizon by either falling through the horizon --- producing a non-vanishing flux $T_{VV}$ --- or by causing a change in the long-range gravitational field $E_{rr}$ on the horizon. Thus, in both the electromagnetic and gravitational cases, memory can be produced on any cosmological horizon. 

To investigate the production of soft gravitons, we consider a quantized gravitational field coupled a classical source --- the stress energy tensor operator is of the form $\op{T}_{ab}=T_{ab}\op{1}$. The expected number of gravitons emitted by $T_{ab}$ corresponds to the norm of the advanced-minus-retarded solution $(ET)_{ab}$. By a similar analysis, any stress energy that causes a change in the long-range gravitational field $E_{rr}$ on the horizon will produce a growing number of soft gravitons. Since the stress energy tensor is conserved one cannot simply create or destroy masses to produce a simple example. However one could consider simply displacing the mass away from an inertial trajectory (see \cite{Danielson:2022sga} for a detailed analysis of this case) or the collision of two inertial masses\footnote{While, in the gravitational case, the source cannot be ``turned off'' at late times one can still calculate the expected number of gravitons emitted using the methods of \cite{Danielson:2025iud}.}.

{It is instructive to compare the above analysis to the analogous story in asymptotically flat spacetimes. There, sources radiate particles to null infinity, and there are relations between the electromagnetic and gravitational memory at infinity analogous to eqs.~\eqref{eq:memeqhorj} and \eqref{eq:bianchi} respectively. It has been shown that, generically, there are inequivalent Hilbert spaces corresponding to different memory sectors at infinity that must be considered even in standard scattering processes (see, e.g.,~\cite{Bloch:1937pw, asymp-quant,Strominger:2017zoo,PSW-IR,memory-orbits,Prabhu:2024lmg}). This situation is under more control than the de Sitter case we study because there is decoupling between the radiation at null infinity and the charged sources which go to timelike infinity, enabling an independent quantization of the charged sources from the quantization of the radiative modes at null infinity.}

{While we have demonstrated in de Sitter that a large and possibly infinite number of soft particles can be produced in the presence of sources, this does not imply that in the fully interacting theory there exist unitarily inequivalent Hilbert spaces. This is because the charged sources do not decouple from the radiation at late times. Indeed, we expect, on general grounds, that when properly treated, there will be a single Hilbert space representation of the interacting field algebra. This expectation comes from an argument that holds in general spacetimes with compact Cauchy surfaces \cite{Wald_1995,Witten:2021jzq} that we summarize now. When quantizing a quantum field, we must deal with an infinite number of modes. For the infinite number of modes with energy larger than the radius of curvature, these may be separated into positive and negative frequency with minimal error and be quantized the same as modes in Minkowski space. Because the spacetime is compact, we are then left with only a finite number of long wavelength modes, which admit a unique representation (up to unitary equivalence) by the Stone-von Neumann theorem. As a simple example, for the massless scalar field, the infrared mode was the constant mode on the three-sphere with unique Hilbert space representation $L^{2}(\mathbb{R})$, which we quantized in sec.~\ref{sec:HilbmmKG} (see remark \ref{rem:DI-TP}). Clearly, this example demonstrates that the existence of a unique Hilbert space does not imply the existence of a normalizable de Sitter invariant state. It would be interesting to understand how these arguments manifest themselves explicitly in fully interacting quantum field theories as well as perturbative quantum gravity.

\begin{acknowledgments}
We would like to thank Dionysios Anninos, Juan Maldacena, Zimo Sun, Robert M.~Wald, and Edward Witten for helpful discussions. JKF is supported by the Marvin L.~Goldberger Member Fund at the Institute for Advanced Study and the National Science Foundation under Grant No. PHY-2207584. GS is supported by the Princeton Gravity Initiative at Princeton University.
\end{acknowledgments}

%%==================================================================================
\appendix 
\section{Covariant construction of the Hilbert space \(\Hilb_\dS\)}
\label{sec:covariant-stuff}

In the main body of the paper we made use of the global coordinates to define the ``pure charge'' mode \(\xi(t)\). We used this mode to define a operator \(\op p(\xi)\), corresponding the constant mode, which acts as a derivative operator on the Hilbert space \(\Hilb_\dS\). However, as mentioned before, the mode \(\xi(t)\) is not singled out in any covariant manner. For instance, any solution \(\gamma(y) = \xi(t) + \Phi(y)\) --- where \(\Phi(y)\) is any solution which has zero charge --- also has unit charge just like \(\xi(t)\). In this appendix, we show that our construction can be carried out completely covariantly without any reference to a choice of ``pure charge'' mode \(\xi(t)\).

Let \(\gamma(y)\) be \emph{any} solution which has unit charge, i.e.,
\be
    q[\gamma] = -\Omega_\Sigma(\gamma, 1) = 1
\ee
Since the charge is conserved, the choice of Cauchy surface used to evaluate the above charge is immaterial. Then, we can define the operator
\be
    \op p(\gamma) \defn - \Omega_\Sigma(\op\phi,\gamma)
\ee
which satisfies the same commutation relation as \cref{eq:commpq}. Thus, we can equivalently consider \(\op p(\gamma)\) as the canonical conjugate operator to the charge operator \(\op q\) for any choice of \(\gamma(y)\).

The construction of the Hilbert space \(\Hilb_\dS\) as a direct integral over the charge value proceeds in exactly the same manner as described above. However, there is no preferred choice of vacuum in any charged Fock space \(\Fock_q\) that can be uniquely specified in a covariant manner. Thus, we must allow for an arbitrary choice of a family of solutions \(\Xi(y;q) = \Xi_q(y)\) with charge \(q\); without loss of generality we can choose this family to be smooth in \(q\). This gives us a non-unique choice of a family of vacua \(\ket{\Omega_{\Xi(q)}} \in \Hilb_\dS\).

Now, for any solution \(\gamma(y)\) of unit charge we define the action of the unitary \(e^{i s \op p(\gamma)}\) on \(\Hilb_\dS\). As before this operator generates a shift of the charge value by \(s\). Under this shift, we get a new family \(\Xi'(y;q)\) of charged solutions
\be
    \Xi'(y;q) &\defn \Xi(y;q-s) + s\gamma(y)
\ee
and a corresponding new family of vacua \(\ket{\Omega_{\Xi'(q)}} \in \Hilb_\dS\). Proceeding as before, we find the action of \(\op p(\gamma)\) on a state in \(\Hilb_\dS\) as
\be
    &\op p(\gamma) \ket{\psi(q)} = \sum_{n=0}^\infty~ \int_{M^n} d^4y_1 \cdots d^4y_n \\
    &\qquad \times \bigg[ i\frac{\partial}{\partial q}\psi(y_1,\ldots,y_n;q) \op\phi(y_1) \cdots \op\phi(y_n) \ket{\Omega_{\Xi(q)}} \\
    &\qquad + \psi(y_1,\ldots,y_n;q) \op\phi(y_1) \cdots \op\phi(y_n) \op\phi(\gamma') \ket{\Omega_{\Xi(q)}} \bigg]
\ee
where
\be
    \gamma'(y;q) \defn - \frac{\partial}{\partial q} \Xi(y; q) + \gamma(y) \,.
\ee
is a function with vanishing charge. This gives a completely covariant definition of the operator \(\op p\) without making any choice of foliation or the corresponding choice of ``pure charge'' mode.

\bibliographystyle{JHEP}
\bibliography{main}
\end{document}

%% file: penrose_ds_tikz.tex
% PENROSE DIAGRAM of a Schwarzschild black hole
\begin{tikzpicture}[scale=3.2]

  \coordinate (-O) at (-1, 0); % center III: origin (r,t) = (0,0)
  \coordinate (-N) at (-1, 1); % north III: t=+infty, i+
  \coordinate (O)  at ( 1, 0); % center I: origin (r,t) = (0,0)
  \coordinate (S)  at ( 1,-1); % south I: t=-infty, i-
  \coordinate (N)  at ( 1, 1); % north I: t=+infty, i+
  \coordinate (E)  at ( 2, 0); % east I:  r=-infty, i0
  \coordinate (W)  at ( 0, 0); % west I:  r=+infty, i0
  \coordinate (B)  at ( 0,-1); % singularity bottom
  
    % \fill[mylightpurple] (-N) -- (N) -- (W) -- cycle;
    % \fill[mylightpurple] (-1,-1) -- (0,0) -- (1,-1) -- cycle;
    % \fill[mylightblue] (N)-- (S) -- (W) -- cycle;
    % \fill[mylightblue] (0,0)-- (-1,-1) -- (-1,1) -- cycle;

  % \node[fill=mylightblue,inner sep=2] at (-.6,-.0) {$\mc{L}$};
  % \node[fill=mylightblue,inner sep=2] at (.6,-.0) {$\mc{R}$};
  % \node[fill=mylightpurple,inner sep=2] at (0,0.6) {$\mc{F}$};
  % \node[fill=mylightpurple,inner sep=2] at (0,-0.7) {$\mc{P}$};

\filldraw[gray, opacity = .2] (N) -- (S) -- (0,0)--cycle;
  % BOUNDARIES
  \draw[thick,black] (N)  -- (S) ;
  \draw[thick,black] (-1,-1)--(-1,1);
  \draw[thick,black] (-1,-1)--(N);
  \draw[thick,black,double] (-1,1) -- (1,1);
  \draw[thick,black,double] (-1,-1) -- (1,-1);

  \draw[thick,dashed,black] (-1,-0.6) to[out=0,in=-180] (1,-0.6);
 % \draw[thick,blue] (-1,1) to[out=-25,in=-180] (1,.2);
 % \draw[thick,green] (0,0) to[out=0,in=-180] (1,.2);
    
   % \node[black,below=-2.5,rotate=-45] at (.5,-0.5) {$\mathcal{H}^-_R$};
    % \node[black,above=-2.5,rotate=-45] at (-.5,0.5) {$\mathcal{H}^-_L$};
        % \node[black,below=-2.5,rotate=45] at (-.5,-0.5) {$\mathcal{H}^+_L$};
    \node[black,above=1,rotate=45] at (0,0) {$\hor \equiv (\eta = \chi)$};
  %   \node[right=1,below right=1,mydarkpurple] at (S) {$i_R^-$};
  % \node[right=1,below left=1,mydarkpurple] at (-1,-1) {$i_L^-$};
  % \node[right=1,above right=1,mydarkpurple] at (1,1) {$i_R^+$};
  % \node[left=1,above left=1,mydarkpurple] at (-1,1) {$i_L^+$};
    \node[above=1,black] at (0,1) {$\scri^+ \equiv (\eta = \pi)$};
    \node[below=1,black] at (0,-1) {$\scri^- \equiv (\eta = 0)$};

    \node[above=1,black] at (1.1,1) {$i^+$};
    \node[below=1,black] at (-1.1,-1) {$i^-$};

    \node[above=1,black, rotate=90] at (-1,0) {$(\chi = 0 )$};
    \node[below=1,black, rotate=90] at (1,0) {$(\chi = \pi )$};

  \filldraw[black] (1,-1) circle (.5pt);
  \filldraw[black] (1,1) circle (.5pt);
  \filldraw[black] (-1,1) circle (.5pt);
  \filldraw[black] (-1,-1) circle (.5pt);
  % \filldraw[black] (0,0) circle (.75pt);

\end{tikzpicture}